\documentclass[onecolumn,english,aps,prb,amsmath,showpacs]{revtex4}
\usepackage{graphicx}
\usepackage{amssymb}

\newcommand{\bvec}[1]{{\mathbf #1}}

\newcommand{\ket}[1]{\left| #1 \right>}

\newcommand{\bq}{\mathbf{q}}
\newcommand{\fx}[2]{#1 \left( #2 \right)}

\begin{document}

\title{Tuning the effects of Landau-level mixing on anisotropic transport in quantum Hall systems}

\author{Peter M. Smith and Malcolm P. Kennett}

\address{Physics Department, Simon Fraser University, 8888 University Drive,
Burnaby, BC, Canada, V5A 1S6}

\date{\today}
\begin{abstract}
Electron-electron interactions in half-filled high Landau levels in
two-dimensional electron gases in a strong perpendicular magnetic
field can lead to states with anisotropic longitudinal resistance.  This longitudinal
resitance is generally believed to arise from broken rotational invariance,
which is indicated by charge density wave (CDW) order in Hartree-Fock 
calculations.  We use the
Hartree-Fock approximation to study the influence of externally tuned
Landau level mixing on the formation of interaction induced states that
break rotational invariance
in two-dimensional electron and hole systems. We focus on the situation
when there are two non-interacting states in the vicinity of
the Fermi level and construct a Landau theory to study coupled
charge density wave order 
that can occur as interactions are tuned and the filling or mixing
are varied. We examine in detail a specific example where mixing is
tuned externally through Rashba spin-orbit coupling. We calculate the
phase diagram and find the possibility of ordering involving
coupled striped or triangular charge density waves in the two levels.
Our results may be relevant to recent transport experiments on quantum
Hall nematics in which Landau-level mixing plays an important role. 
\end{abstract}


\maketitle

\section{Introduction}

The integer\cite{VonKlitzing} and fractional\cite{Stormer} quantum
Hall effects have been studied extensively in two-dimensional electron
systems (2DES).\cite{PrangeGirvin} It has been proposed\cite{HLR,MooreRead}
and observed\cite{Willett} that the states at half odd integer filling
fraction in low Landau levels (LLs) have distinct behaviour from states
at odd denominator filling fractions. This distinction extends to
higher LLs, where anisotropic transport has been observed at half
odd integer filling fractions,\cite{Lilly,Pan,Lilly2} which is believed to arise
from broken rotational symmetry due to electron-electron
interactions.\cite{FPA,MoessnerChalker,KFS,FKS,Fogler,FradkinKivelson,Yang,MacDonaldFisher,FKMN,Dorsey1,Dorsey2,Barci,Doan,FradkinReview} 
The experimental
evidence for broken rotation symmetry at half odd 
integer filling fractions is large peaks in the longitudinal resistivity, 
which vanish as the
direction of applied current is rotated by ninety degrees. 

Hartree-Fock calculations of the effects of electron-electron interactions in
high Landau levels indicated a transition to a striped charge density wave (CDW) 
phase at half odd integer filling fractions.\cite{MoessnerChalker,KFS,FKS,Fogler}
It has been argued by Fradkin and Kivelson and others that
the anisotropy is evidence of an electronic nematic 
phase.\cite{FradkinKivelson,FKMN,Dorsey1,Dorsey2,Barci,Doan,FradkinReview}
However there continues to be debate as to whether the 
states are electronic nematics or smectics.\cite{FradkinReview, recentTsui, Ciftja} 

There have been several recent experiments on two-dimensional
electron or hole systems that are in or close to the regime where
anisotropic quantum Hall states are 
expected for which Landau level mixing is believed to be either 
appreciable or tunable via an external parameter.\cite{Manfra,Gusev,Grayson}
LL mixing can arise from a variety of sources, such as disorder, geometric
confinement effects in a quantum well, interactions or spin-orbit
coupling\cite{WinklerBook}  but has generally been 
ignored in theoretical studies of anisotropic quantum Hall 
states.\cite{FradkinKivelson,KFS,MoessnerChalker,FPA}
A couple of exceptions are investigations of 
the effects of LL mixing from electron-electron
interactions\cite{MacDonald} or disorder\cite{Stanescu} on CDW formation 
which may be relevant for experiments showing re-entrant behaviour
of quantum Hall states in the lowest LL.\cite{Csathy} There has also
been recent interest in LL mixing in the context of non-Abelian quantum Hall
states.\cite{Simon,Nayak,Chesi}

We investigate the effects of Landau level mixing on the
formation of quantum Hall states with broken rotation symmetry
by using  the Hartree-Fock approximation to study CDW
formation in a two-dimensional system of charged fermions (electrons
or holes) when LL mixing and interactions are present.
 This neglects quantum and thermal fluctuations,\cite{Dorsey1}
but can be hoped to be as informative as Hartree-Fock studies on
single Landau levels.\cite{MoessnerChalker,KFS}  We assume that
there is a term in the Hamiltonian that mixes LLs, leading to  
non-interacting single-particle eigenstates that are 
linear combinations of LLs. We explore
the effects of interactions on CDW ordering within these eigenstates.
We specialize to the situation when two energy levels lie
close to the Fermi energy, $E_{F}$, of the system, which can lead
to competition between CDW phases of various symmetries originating
in different non-interacting states. To study this, we construct
a Landau theory for CDW ordering in the presence of LL mixing
when there are two states that are close to the Fermi energy.  
In the range of energies where the two states are close to degenerate
we also consider the possibility of quantum Hall ferromagnetism, but for 
the example we consider here, CDW ordering occurs at higher temperatures
within the Hartree-Fock approximation.

In order to study the effects of LL mixing on CDW formation 
experimentally in a systematic way, some external tuning parameter that
 can mix LLs is required. Neither disorder nor interactions
are particularly suitable for this task. However, Rashba spin-orbit coupling can
be tuned experimentally in a quantum well by parameters such as applied
gate voltage. This motivates us to investigate the phase diagram as a function
of filling fraction, LL mixing and temperature when LLs are mixed
via Rashba spin-orbit coupling when two states are relatively close to the
Fermi level.  We find that at half-integer filling fractions there can be
triangular CDW ordering in {\it both} levels simultaneously
when the filling of the higher energy level is non-zero.

This paper is organized as follows. In Sec.~\ref{sec:Moessner} we
investigate electron-electron scattering in the presence of LL mixing
in high LLs, using a similar approach to Moessner and Chalker,\cite{MoessnerChalker}
to argue that there is still an instability towards CDW ordering in the presence of 
LL mixing. In Sec.~\ref{sec:Landau-free-energy-theory} we construct a Landau theory
for striped and triangular CDW order when there are two states close
to the Fermi energy.
In Sec.~\ref{sec:Numerics} we display numerically calculated phase
diagrams for LL mixing due to Rashba spin orbit coupling and also
discuss possible quantum Hall ferromagnetism near degeneracy.  
We discuss our results and their possible connection to experiment in 
Sec.~\ref{sec:Conclusion}.

\section{Interactions in the presence of Landau level
mixing}

\label{sec:Moessner}

We consider fermions (electrons or holes) confined to two dimensions
in a perpendicular magnetic field. We label single-particle basis
states as $\left|nkj_{z}\right>$, where $n$ labels LL index, $k$
is the pseudomomentum, and $j_{z}$ is the spin eigenvalue. The projection
of the orbital part of the eigenstates onto real-space is the LL wavefunction
\begin{eqnarray}
\psi_{k}^{\left(n\right)}\left(\mathbf{r}\right) & = & \frac{1}{\sqrt{2^{n}n!\sqrt{\pi}l_{0}L}}e^{iky}\phi_{n}\left(\frac{x-kl_{0}^{2}}{l_{0}}\right),
\label{eq:LLwavefn}
\end{eqnarray}
 where we work in the Landau gauge $\mathbf{A}\left(\mathbf{r}\right)=B\left(-y,0,0\right)$,
with $L$ the spatial extent of the system and $l_{0}=\sqrt{h/eB}$
the magnetic length.

We assume that there is some source of mixing of the basis states,
such as spin-orbit coupling, which leads to a set of non-interacting
single-particle eigenstates of the form \begin{eqnarray}
\left|\Phi k\right> & = & \sum_{\alpha=\{n_{\alpha},j_{z}^{\alpha}\}}C_{\Phi\alpha}\left|\alpha k\right\rangle ,\label{eq:Eigenstates}\end{eqnarray}
 where the $C_{\Phi\alpha}$ are the coefficients of the basis states
and $\left|\alpha k\right\rangle $ is shorthand for $\left|n_{\alpha}kj_{z}^{\alpha}\right\rangle $.

We now allow for interactions between fermions via a potential $V\left(\mathbf{r}\right)$,
which has Fourier transform $\tilde{V}\left(\mathbf{q}\right)$, leading
to the interaction Hamiltonian 
 \begin{eqnarray}
H_{I} 
& = & \frac{1}{2}\sum_{\left\{ \Phi_{1},\Phi_{2},\Phi_{3},\Phi_{4}\right\} }\sum_{\left\{ k_{1},k_{2},k_{3},k_{4}\right\} }a_{\Phi_{1}k_{1}}^{\dagger}a_{\Phi_{2}k_{2}}^{\dagger}a_{\Phi_{3}k_{3}}a_{\Phi_{4}k_{4}} \nonumber\\
&     &\hspace{1.5cm}\times\int\frac{\mbox{d}^{2}\mathbf{q}}{\left(2\pi\right)^{2}}\tilde{V}\left(\mathbf{q}\right)\left\langle \Phi_{1}k_{1};\Phi_{2}k_{2}\left|e^{i\mathbf{q}\cdot\left(\mathbf{r_{2}-r_{1}}\right)}\right|\Phi_{3}k_{3};\Phi_{4}k_{4}\right\rangle \nonumber \\
& = & \frac{1}{2}\sum_{\left\{ \Phi_{1},\Phi_{2},\Phi_{3},\Phi_{4}\right\} }\sum_{\left\{ k_{1},k_{2},k_{3},k_{4}\right\} }a_{\Phi_{1}k_{1}}^{\dagger}a_{\Phi_{2}k_{2}}^{\dagger}a_{\Phi_{3}k_{3}}a_{\Phi_{4}k_{4}}\sum_{\alpha\beta\sigma\rho}C_{\Phi_{1}\alpha}^{*}C_{\Phi_{2}\sigma}^{*}C_{\Phi_{3}\rho}C_{\Phi_{4}\beta}\delta_{j_z^{\alpha}-j_z^{\beta}+j_z^{\gamma}-j_z^{\delta}}\nonumber \\
&  & \hspace{1.5cm}\times\int\frac{\mbox{d}^{2}\mathbf{q}}{\left(2\pi\right)^{2}}\tilde{V}\left(\mathbf{q}\right)\left\langle n_{\alpha}k_{1}\left|e^{-i\mathbf{q}\cdot\mathbf{r_{1}}}\right|n_{\beta}k_{4}\right\rangle \left\langle n_{\sigma}k_{2}\left|e^{i\mathbf{q}\cdot\mathbf{r_{2}}}\right|n_{\rho}k_{3}\right\rangle ,\label{eq:InteractionHamiltonian1}\end{eqnarray}
 where $a_{\Phi k}$ is the annihilation operator of $\ket{\Phi k}$,
and the four single particle states involved in the interaction are
labelled $\left\{ \Phi_{1},\Phi_{2},\Phi_{3},\Phi_{4}\right\} $ with
corresponding pseudomomenta $\left\{ k_{1},k_{2},k_{3},k_{4}\right\} $.
The orbital matrix elements have been evaluated by Raikh and Shahbazyan\cite{Raikh} 
to be 
\begin{eqnarray}
\left\langle n_{\alpha}k_{1}\left|e^{-i\mathbf{q}\cdot\mathbf{r_{1}}}\right|n_{\beta}k_{2}\right\rangle  
& = & \frac{2\pi}{L}\delta\left(k_{1}-k_{2}-q_{y}\right)A_{n_{\alpha}n_{\beta}}\left(\frac{1}{2}q^{2}l_{0}^{2}\right)e^{i\left[\frac{1}{2}l_{0}^{2}q_{x}(k_{1}+k_{2})+\left(n_{\alpha}-n_{\beta}\right)\left(\theta-\frac{\pi}{2}\right)\right]},\nonumber \\
&    &
\end{eqnarray}
 where 
 \begin{eqnarray*}
A_{n_{\alpha}n_{\beta}}\left(x\right) & = & \left(\frac{n_{\alpha\beta}^{'}!}{n_{\alpha\beta}!}\right)^{\frac{1}{2}}x^{\frac{\Delta n_{\alpha\beta}}{2}}e^{-\frac{x}{2}}\mathcal{L}_{n_{\alpha\beta}^{'}}^{\Delta n_{\alpha\beta}}\left(x\right),\end{eqnarray*}
 with $n_{\alpha\beta}=\max\left(n_{\alpha},n_{\beta}\right)$, $n_{\alpha\beta}^{'}=\min\left(n_{\alpha},n_{\beta}\right)$,
$\Delta n_{\alpha\beta}=\left|n_{\alpha}-n_{\beta}\right|$, and $\theta=\arctan\left(q_{y}/q_{x}\right)$.
Equation~(\ref{eq:InteractionHamiltonian1}) can be rewritten
as 
\begin{eqnarray}
H_{I} & = & \frac{1}{2}\sum_{\{\Phi_{1},\Phi_{2},\Phi_{3},\Phi_{4}\}}\sum_{klm}M_{\{\Phi_{1},\Phi_{2},\Phi_{3},\Phi_{4}\}}\left(m,l\right)a_{\Phi_{1}k+l}^{\dagger}a_{\Phi_{2}k+m}^{\mbox{\mbox{ }}\dagger}a_{\Phi_{3}k+l+m}^{\mbox{ }}a_{\Phi_{4}k}^{\mbox{ }},
\end{eqnarray}
 where 
 \begin{eqnarray}
M_{\{\Phi_{1},\Phi_{2},\Phi_{3},\Phi_{4}\}}\left(m,l\right) & = & \sum_{\alpha\beta\sigma\rho}C_{\Phi_{1}\alpha}^{*}C_{\Phi_{2}\sigma}^{*}C_{\Phi_{3}\rho}C_{\Phi_{4}\beta}\mathbf{}\delta_{\alpha\beta}\delta_{\sigma\rho}I_{\alpha\sigma\rho\beta}\left(m,l\right),\label{eq:InteractionVertexEigenstates}
\end{eqnarray}
 and 
 \begin{eqnarray}
I_{\alpha\sigma\rho\beta}\left(m,l\right) & = & \frac{2\pi}{L}\int\frac{\mbox{d}^{2}\mathbf{q}}{\left(2\pi\right)^{2}}\tilde{V}\left(q\right)\delta\left(q_{y}-l\right)A_{n_{\alpha}n_{\beta}}\left(\frac{1}{2}q^{2}l_{0}^{2}\right)A_{n_{\sigma}n_{\rho}}\left(\frac{1}{2}q^{2}l_{0}^{2}\right) \nonumber \\
    &    & \hspace{1.5cm} \times e^{-il_{0}^{2}q_{x}m}e^{i\left(\theta-\frac{\pi}{2}\right)\left(n_{\alpha}-n_{\beta}+n_{\sigma}-n_{\rho}\right)}.\label{eq:OrbitalInteraction}
    \end{eqnarray}

\begin{figure}[ht]
 \includegraphics[scale=0.6]{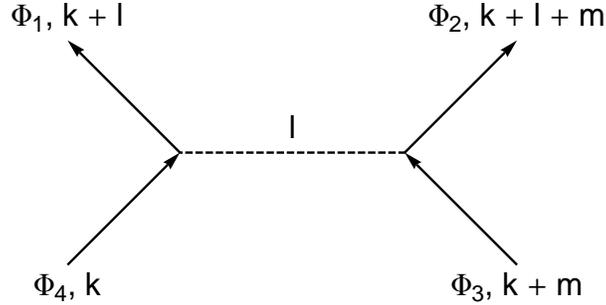}
\caption{Interaction between states $\Phi_{1}$, $\Phi_{2}$, $\Phi_{3}$,
and $\Phi_{4}$. $k$, $l$, and $m$ are pseudomomenta.}
\label{fig:OneParticleVertex} 
\end{figure}

The function $M_{\{\Phi_{1},\Phi_{2},\Phi_{3},\Phi_{4}\}}\left(m,l\right)$
describes the strength of the interaction between the eigenstates
$\Phi_{1},\Phi_{2},\Phi_{3},$ and $\Phi_{4}$ (illustrated in Fig.~\ref{fig:OneParticleVertex}),
with the function $I_{\alpha\sigma\rho\beta}\left(m,l\right)$ describing
the overlap between the orbital part of the basis states. The strength
of the interactions between particular eigenstates can be tuned by
using external parameters to vary the coefficients, $C_{\Phi_{i}\alpha}$.

\subsection{Hartree-Fock Approximation\label{sec:2DHSs-in-Large-LL}}

We now investigate the self-energy and two particle scattering vertex
using the Hartree-Fock approximation. Moessner and Chalker \cite{MoessnerChalker}
showed that for fermions interacting with the hard-core potential
$\tilde{V}(q)=-vq^{2}$ within a single LL, the Hartree-Fock approximation
is exact in the limit of large LL index. We generalize this result
to electrons with different LL indices interacting via a hard-core
interaction. Provided the difference in LL index is much less than
the maximum LL index, then in the limit that the maximum LL index
is large, the Hartree-Fock approximation is asymptotically exact for
the hard-core potential. We return to consider Coulomb interactions
in Sec.~\ref{sec:Landau-free-energy-theory}.

To take the large-LL limit, we follow Moessner and Chalker and replace
$A_{n_{\alpha}n_{\beta}}(x)$ by its WKB envelope: 
\begin{eqnarray}
A_{n_{\alpha}n_{\beta}}\left(x\right) & = & \frac{1}{\sqrt{\pi}}\left[\frac{n'_{\alpha\beta}!}{\left(n'_{\alpha\beta}+\Delta n_{\alpha\beta}\right)!}\right]^{1/2} 
\frac{1}{\left[4n'_{\alpha\beta}x-\left(x-\Delta n_{\alpha\beta}\right)^{2}\right]^{1/4}}. 
\label{eq:WKBEnvelope}
\end{eqnarray}
The envelope is defined for $x_{-}^{\alpha\beta}<x<x_{+}^{\alpha\beta}$,
where \begin{eqnarray*}
x_{\pm}^{\alpha\beta} & = & \left(2n_{\alpha\beta}'+\Delta n_{\alpha\beta}\right)\left[1\pm\sqrt{1-\left(\frac{\Delta n_{\alpha\beta}}{\Delta n_{\alpha\beta}+2n_{\alpha\beta}'}\right)}\right].\end{eqnarray*}

We now turn to the properties of the interaction vertex $M_{\{\Phi_{1},\Phi_{2},\Phi_{3},\Phi_{4}\}}$
in the limit of large LLs; specifically, we study the properties of
function $I_{\alpha\sigma\rho\beta}\left(m,l\right)$ {[}Eq.~(\ref{eq:OrbitalInteraction}){]}.
The function $I_{\alpha\sigma\rho\beta}\left(m,l\right)$ gives the
strength of the interaction between eigenstates in terms of the constituent
basis states. For the single-particle and two-particle
vertices, we focus on the situation when there are two eigenstates
lying near the Fermi level.

Substitution of Eq.~(\ref{eq:WKBEnvelope}) into Eq.~(\ref{eq:OrbitalInteraction})
yields
 \begin{eqnarray}
I_{\alpha\sigma\rho\beta}\left(m,l\right) 
& = & \frac{4}{Ll_{0}}\left[\frac{n'_{\alpha\beta}!}{\left(n'_{\alpha\beta}+\Delta n_{\alpha\beta}\right)!}\right]^{1/2}\left[\frac{n'_{\sigma\rho}!}{\left(n'_{\sigma\rho}+\Delta n_{\sigma\rho}\right)!}\right]^{1/2}\int_{q_{x}^{{\rm min}}}^{q_{x}^{{\rm max}}}\frac{\mbox{d}q_{x}}{2\pi}\nonumber \\
 &    & \times\frac{\tilde{V}\left(q_{x},l\right)\cos\left[l_{0}^{2}q_{x}m-\left(n_{\alpha}-n_{\beta}+n_{\sigma}-n_{\rho}\right)\left(\arctan\left(\frac{l}{q_{x}}\right)-\frac{\pi}{2}\right)\right]}{\left[8n'_{\alpha\beta}\left(q_{x}^{2}+l^{2}\right)-\left(q_{x}^{2}+l^{2}-2\frac{\Delta n_{\alpha\beta}}{l_{0}^{2}}\right)^{2}\right]^{1/4}\left[8n'_{\rho\sigma}\left(q_{x}^{2}+l^{2}\right)-\left(q_{x}^{2}+l^{2}-2\frac{\Delta n_{\rho\sigma}}{l_{0}^{2}}\right)^{2}\right]^{1/4}}.\nonumber \\
 &     &\label{eq:InteractionWKB1}
 \end{eqnarray}
 As the eigenstates in the presence of LL mixing are
not Landau levels, when we refer to the large-LL limit, we mean that
the non-interacting single-particle eigenstates near the Fermi energy
are dominated by contributions from Landau levels with large indices.
If $\Delta n_{\alpha\beta}\neq0$, and $\Delta n_{\alpha\beta}\ll n'_{\alpha\beta}$,
then $$
\sqrt{\frac{n'_{\alpha\beta}!}{(n'_{\alpha\beta}+\Delta n_{\alpha\beta})!}}\sim\left(\frac{1}{n'_{\alpha\beta}}\right)^{\frac{\Delta n_{\alpha\beta}}{2}}\to0$$
as $n'_{\alpha\beta}\to\infty$, which causes the interaction
vertex $I_{\alpha\sigma\rho\beta}\left(m,l\right)$ to vanish. Thus,
in this limit, $I_{\alpha\sigma\rho\beta}\left(m,l\right)$ is finite
only when $\Delta n_{\alpha\beta}=\Delta n_{\sigma\rho}=0$, in which
case \begin{eqnarray}
I_{\alpha\sigma\rho\beta}\left(m,l\right) & = & \frac{4}{Ll_{0}}\delta_{n_{\alpha},n_{\beta}}\delta_{n_{\sigma},n_{\rho}}\int_{0}^{q_{x}^{{\rm max}}}\frac{dq_{x}}{2\pi}\frac{\tilde{V}\left(q_{x},l\right)\cos\left[l_{0}^{2}q_{x}m\right]}{\sqrt{q_{x}^{2}+l^{2}}\left[8n{}_{\alpha}-\left(q_{x}^{2}+l^{2}\right)\right]^{1/4}\left[8n_{\sigma}-\left(q_{x}^{2}+l^{2}\right)\right]^{1/4}}, \nonumber \\
& &
\label{eq:Ideltanzero}
\end{eqnarray}
 where \begin{eqnarray*}
q_{x}^{{\rm max}} & = & \sqrt{\left(q^{{\rm max}}\right)^{2}-l^{2}},\\
q^{{\rm max}} & = & {\rm \max}\left\{ \sqrt{8n_{a}},\sqrt{8n_{\sigma}}\right\} .\end{eqnarray*}
 If $n_{\alpha}=n_{\sigma}$, the envelopes in Eq.~(\ref{eq:Ideltanzero})
associated with $n_{\alpha}$ and $n_{\sigma}$ overlap exactly, whereas
when $n_{\alpha}\ne n_{\sigma}$ there is less than perfect overlap,
so $I_{\alpha\alpha\alpha\alpha}(m,l)$ will be an upper bound on
the interaction vertex.

Let ${\mathcal{N}}$ denote the index of the highest LL basis state
that contributes to occupied non-interacting eigenstates. Moessner
and Chalker showed that for a contact potential $I_{\alpha\alpha\alpha\alpha}(m,l)$
is non-zero only when $m=0$ in the large-LL limit, hence, as we have
established that $I_{\alpha\alpha\alpha\alpha}(m,l)$ is an upper
bound for $I_{\alpha\sigma\rho\beta}\left(m,l\right)$, it follows
that $I_{\alpha\sigma\rho\beta}\left(m,l\right)$ is nonzero only
when $m=0$, and obeys the bound \begin{eqnarray*}
-\sum_{q_{y}}I_{\alpha\sigma\rho\beta}\left(0,q_{y}\right) & \le & \frac{2v{\mathcal{N}}}{\pi}.\label{eq:BoundOnInteractions}\end{eqnarray*}
 Thus, in the large-LL limit, only exchange terms are important in
the interactions between basis states for a contact potential. The
results of MC for other classes of diagrams can also be applied here
as an upper bound to other types of diagrammatic contributions. Thus,
diagrams with crossed interaction lines and closed fermion loops vanish
in the large-LL limit. When we consider the case where the relevant
constituent LL indices are not necessarily large, we apply the Hartree-Fock
approximation directly. In this situation, both Hartree terms and
Fock terms are important.

To avoid divergence of the bound on $I_{\alpha\sigma\rho\beta}$ as
the number of constituent basis states increases, we scale $v$ with
$\mathcal{N}$ such that $v\mathcal{N}$ is constant. Assuming that
each eigenstate $\Phi$ is a linear combination of $\mathcal{P}$
LL basis states (for example for Rashba spin-orbit coupling considered
in Sec.~\ref{sec:Numerics}, ${\mathcal{P}}=4$), we obtain the following
upper bound on the interaction vertex: \begin{eqnarray*}
-\sum_{l}M_{\{\Phi_{1},\Phi_{2},\Phi_{3},\Phi_{4}\}}\left(0,l\right) & \le & \frac{2v\mathcal{P}\mathcal{N}}{\pi}.\end{eqnarray*}

\subsubsection{Single-particle propagator: Self-energy}

We now turn to consider the self-energy in the presence of LL mixing.
The finite-temperature Green's function for an eigenstate $\Phi$
is \begin{eqnarray*}
G\left(\Phi,\omega_{n}\right) & = & \frac{1}{i\omega_{n}-\xi_{\Phi}-\Sigma_{\Phi}},\end{eqnarray*}
 where $\omega_{n}=\left(2n+1\right)\pi/\beta$ is the $n^{{\rm th}}$
Matsubara frequency, $\xi_{\Phi}=E_{\Phi}-\mu$, where $E_{\Phi}$
is the energy of non-interacting eigenstate $\Phi$, $\mu$ is the
chemical potential, and $\Sigma_{\Phi}$ is the self-energy. We note
that the bare Green's function, $G_{0}\left(\Phi,\omega_{n}\right)=\left(i\omega_{n}-\xi_{\Phi}\right)^{-1}$
is independent of pseudomomentum $k$: the LL mixing is chosen so that the energy is independent of $k$, as for single LLs.

In this section we determine $\Sigma_{\Phi}$ using the Dyson equation,
keeping only terms which do not vanish in the large-LL limit. Note
that all Hartree diagrams vanish for the special choice of a contact
potential, so we need only consider the Fock diagrams. For a diagram
with a single propagator and a single interaction line, the summations
over Matsubara frequencies and momenta $l$ decouple. Then \begin{eqnarray}
\Sigma_{\Phi} & = & -\frac{1}{\beta}\sum_{\Phi_{\alpha}}\sum_{j,\omega_{m}}G\left(\Phi_{\alpha},\omega_{m}\right)M_{\Phi\Phi_{\alpha}\Phi\Phi_{\alpha}}\left(0,j\right).\label{eq:self_phi}\end{eqnarray}
 The interaction vertex corresponding to $M_{\Phi\Phi_{\alpha}\Phi\Phi_{\alpha}}\left(0,j\right)$
is illustrated in Fig.~\ref{fig:OneParticleVertex}. Since we consider
only exchange interactions in Eq.~(\ref{eq:self_phi}), we adopt
the notation $M_{\Phi\Phi_{\alpha}\Phi\Phi_{\alpha}}\left(0,j\right)=M_{\Phi\Phi_{\alpha}}\left(0,j\right)$
for brevity.

The full Green's function can be expressed in terms of the bare propagator,
$G_{0}\left(\Phi,\omega_{n}\right)$ as  
\begin{eqnarray*}
G\left(\Phi,\omega_{n}\right) 
& = & G_{0}\left(\Phi,\omega_{n}\right)+\frac{1}{\beta}G_{0}\left(\Phi,\omega_{n}\right)^{2}\sum_{\Phi_{\alpha}}\sum_{j,\omega_{m}}G_{0}\left(\Phi_{\alpha},\omega_{m}\right)M_{\Phi\Phi_{\alpha}}\left(0,j\right)\nonumber\\
&  & +\frac{1}{\beta^{2}}G_{0}\left(\Phi,\omega_{n}\right)^{2}\sum_{\Phi_{\alpha},\Phi_{\beta}}\sum_{j,k}\sum_{\omega_{m},\omega_{p}}G_{0}\left(\Phi_{\alpha},\omega_{m}\right)M_{\Phi\Phi_{\alpha}}\left(0,j\right)G_{0}\left(\Phi_{\beta},\omega_{p}\right)M_{\Phi_{\alpha}\Phi_{\beta}}\left(0,k\right)\nonumber \\
&   &+\ldots\,.
\end{eqnarray*}
 We can simplify each term in the expansion by noting that the sums
over Matsubara frequencies and momenta decouple. Let $\mathcal{M}_{\Phi_{\alpha}\Phi_{\beta}}=\sum_{j}M_{\Phi_{\alpha}\Phi_{\beta}}\left(0,j\right)$.
Then \begin{eqnarray*}
\Sigma_{\Phi} & = & -\frac{1}{\beta}\sum_{\Phi_{\alpha}}\sum_{\omega_{n}}G_{0}\left(\Phi_{\alpha},\omega_{n}\right)\mathcal{M}_{\Phi\Phi_{\alpha}}-\frac{1}{\beta^{2}}\sum_{\Phi_{\alpha},\Phi_{\beta}}\sum_{\omega_{n},\omega_{m}}G_{0}\left(\Phi_{\alpha},\omega_{n}\right)^{2}\mathcal{M}_{\Phi\Phi_{\alpha}}G_{0}\left(\Phi_{\beta},\omega_{m}\right)\mathcal{M}_{\Phi_{\alpha}\Phi_{\beta}}\\
 &  & -\frac{1}{\beta^{3}}\sum_{\Phi_{\alpha},\Phi_{\beta},\Phi_{\gamma}}\sum_{\omega_{n},\omega_{m},\omega_{p}}G_{0}\left(\Phi_{\alpha},\omega_{n}\right)^{2}\mathcal{M}_{\Phi\Phi_{\alpha}}G_{0}\left(\Phi_{\beta},\omega_{m}\right)^{2}\mathcal{M}_{\Phi_{\alpha}\Phi_{\beta}}G_{0}\left(\Phi_{\gamma},\omega_{p}\right)\mathcal{M}_{\Phi_{\beta}\Phi_{\gamma}}.\end{eqnarray*}
Noting $\frac{1}{\beta}\sum_{\omega_{n}}G_{0}\left(\Phi,\omega_{n}\right)$=$f\left(\xi_{\Phi}-\mu\right)$,
$\frac{1}{\beta}\sum_{\omega_{n}}G_{0}\left(\Phi,\omega_{n}\right)^{2}$=$-\beta f\left(\xi_{\Phi}-\mu\right)\left[1-f\left(\xi_{\Phi}-\mu\right)\right]$,
and $\nu_{\Phi}$=$f\left(\xi_{\Phi}-\mu\right)$, we get
 \begin{eqnarray}
\Sigma_{\Phi} & = & -\sum_{\Phi_{\alpha}}\nu_{\Phi_{\alpha}}\mathcal{M}_{\Phi\Phi_{\alpha}}+\beta\sum_{\Phi_{\alpha},\Phi_{\beta}}\nu_{\Phi_{\alpha}}\left(1-\nu_{\Phi_{\alpha}}\right)\mathcal{M}_{\Phi\Phi_{\alpha}}\nu_{\Phi_{\beta}}\mathcal{M}_{\Phi_{\alpha}\Phi_{\beta}}\nonumber \\
 &  & -\beta^{2}\sum_{\Phi_{\alpha},\Phi_{\beta},\Phi_{\gamma}}\nu_{\Phi_{\alpha}}\left(1-\nu_{\Phi_{\alpha}}\right)\mathcal{M}_{\Phi\Phi_{\alpha}}\nu_{\Phi_{\beta}}\left(1-\nu_{\Phi_{\beta}}\right)\mathcal{M}_{\Phi_{\alpha}\Phi_{\beta}}\nu_{\Phi_{\gamma}}\mathcal{M}_{\Phi_{\beta}\Phi_{\gamma}}+\ldots\,.\label{eq:SelfEnergy1}\end{eqnarray}

\subsubsection{Two-particle vertex: CDW Instability}

\label{sub:Two-particle-vertex:-Instabilities} We now turn to investigate
the two-particle vertex, which was shown by Moessner and Chalker to
indicate an instability towards charge density wave (CDW) order in
the case of a single Landau level. We show that such an instability
persists when two levels are involved and identify temperature
scales corresponding to CDW ordering in the specific example when
there are only two states, $\Phi_{1}$ and $\Phi_{2}$, near the Fermi
energy. Moreover, we identify a third temperature scale related to
the possibility of mixed order, which we discuss in more detail in
Sec.~\ref{sec:Landau-free-energy-theory}.

\begin{figure}
\begin{centering}
\includegraphics[scale=0.7]{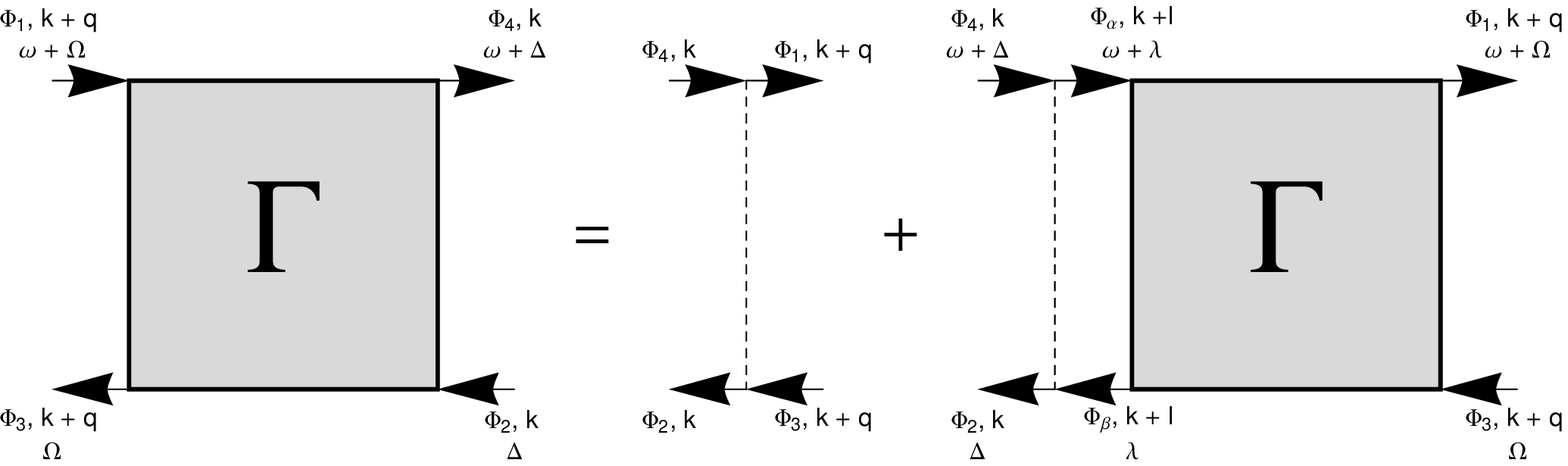} 
\par\end{centering}

\caption{Diagrammatic representation of Bethe-Salpeter equation. \label{fig:BetheSalpeter}}

\end{figure}

Consider the two-particle vertex, which can be determined via the
Bethe-Salpeter equation (Fig.~\ref{fig:BetheSalpeter}): 
\begin{eqnarray}
\Gamma_{\Phi_{\alpha}\Phi_{\beta}\Phi_{\gamma}\Phi_{\delta}}\left(q,\omega\right) 
& = & M_{\Phi_{\alpha}\Phi_{\beta}\Phi_{\gamma}\Phi_{\delta}}\left(0,q\right)\nonumber \\
&    & - \frac{1}{\beta}\sum_{\Phi_{j}\Phi_{k}}\sum_{\lambda,l}M_{\Phi_{k}\Phi_{\beta}\Phi_{j}\Phi_{\delta}}\left(0,l\right)G_{0}\left(\Phi_{k},\omega+\lambda\right)G_{0}\left(\Phi_{j},\lambda\right)\Gamma_{\Phi_{\alpha}\Phi_{j}\Phi_{\gamma}\Phi_{k}}\left(q-l,\omega\right).\nonumber \\
&     &
\end{eqnarray}
 The summations run over all momenta, Matsubara frequencies,
and eigenstates.

First, we examine the case when $\omega\neq0$. We rewrite the sum
over states as two separate sums: a single sum over $\Phi_{j}$ for
the terms where $\Phi_{j}=\Phi_{k}$ along with a sum over $\Phi_{j}\neq\Phi_{k}$.
The first sum vanishes when $\omega\neq0$, whereas for the second
sum, $-\frac{1}{\beta}\sum_{\lambda}G_{0}\left(\Phi_{k},\omega+\lambda\right)G_{0}\left(\Phi_{j},\lambda\right)=\frac{\nu_{\Phi_{k}}-\nu_{\Phi_{j}}}{i\omega-\left(\xi_{\Phi_{k}}-\xi_{\Phi_{j}}\right)}.$
Then, the two-particle vertex is 
\begin{eqnarray}
\Gamma_{\Phi_{\alpha}\Phi_{\beta}\Phi_{\gamma}\Phi_{\delta}}\left(q,\omega\right) 
& = & M_{\Phi_{\alpha}\Phi_{\beta}\Phi_{\gamma}\Phi_{\delta}}\left(0,q\right) \nonumber \\
&    &+\sum_{\Phi_{j}\neq\Phi_{k}}\sum_{l}M_{\Phi_{k}\Phi_{\beta}\Phi_{j}\Phi_{\delta}}\left(0,l\right)\Gamma_{\Phi_{\alpha}\Phi_{j}\Phi_{\gamma}\Phi_{k}}\left(q-l,\omega\right)\frac{\nu_{\Phi_{k}}-\nu_{\Phi_{j}}}{i\omega-\left(\xi_{\Phi_{k}}-\xi_{\Phi_{j}}\right)} .\nonumber \\
&    &
\label{eq:BetheSalpeterOmegaFinite}
\end{eqnarray}

Now, consider $\omega=0$. We separate the summation over eigenstates
into two cases: $\Phi_{j}=\Phi_{k}$ and $\Phi_{j}\neq\Phi_{k}$.
If $\Phi_{j}=\Phi_{k}$, then $\frac{1}{\beta}\sum_{\lambda}G_{0}\left(\Phi_{j},\lambda\right)^{2}=-\beta\nu_{\Phi_{j}}\left(1-\nu_{\Phi_{j}}\right).$
If $\Phi_{j}\neq\Phi_{k}$, then $\frac{1}{\beta}\sum_{\lambda}G_{0}\left(\Phi_{j},\lambda\right)G_{0}\left(\Phi_{k},\lambda\right)=\frac{\nu_{\Phi_{k}}-\nu_{\Phi_{j}}}{\xi_{\Phi_{k}}-\xi_{\Phi_{j}}}.$
This leads to \begin{eqnarray}
\Gamma_{\Phi_{\alpha}\Phi_{\beta}\Phi_{\gamma}\Phi_{\delta}}\left(q,0\right) & = & M_{\Phi_{\alpha}\Phi_{\beta}\Phi_{\gamma}\Phi_{\delta}}\left(0,q\right)+\beta\sum_{\Phi_{j},l}\nu_{\Phi_{j}}\left(1-\nu_{\Phi_{j}}\right)M_{\Phi_{j}\Phi_{\beta}\Phi_{j}\Phi_{\delta}}\left(0,l\right)\Gamma_{\Phi_{\alpha}\Phi_{j}\Phi_{\gamma}\Phi_{j}}\left(q-l,0\right)\nonumber \\
\mbox{ } &  & -\sum_{\Phi_{j},\Phi_{k}}\sum_{l}\frac{\nu_{\Phi_{k}}-\nu_{\Phi_{j}}}{\xi_{\Phi_{k}}-\xi_{\Phi_{j}}}M_{\Phi_{k}\Phi_{\beta}\Phi_{j}\Phi_{\delta}}\left(0,l\right)\Gamma_{\Phi_{\alpha}\Phi_{j}\Phi_{\gamma}\Phi_{k}}\left(q-l,0\right).\label{eq:BetheSalpeterOmegaZero}\end{eqnarray}
 Equations (\ref{eq:BetheSalpeterOmegaFinite}) and
(\ref{eq:BetheSalpeterOmegaZero}) are exact expressions for the two-particle
vertex at finite and zero frequency, respectively.

The expressions in Eqs.~(\ref{eq:BetheSalpeterOmegaFinite}) and (\ref{eq:BetheSalpeterOmegaZero})
are somewhat cumbersome. Hence, it is instructive to investigate a
simpler case than the general situation, in particular, the case where
only two states are located near the Fermi energy.

\subsubsection{Two states near the Fermi energy}

Let $\Phi_{1}$ and $\Phi_{2}$, with energies $\xi_{\Phi_{1}}\leq\xi_{\Phi_{2}}$,
label the two states that lie nearest the Fermi surface and assume
that both $\nu_{\Phi_{1}}$ and $\nu_{\Phi_{2}}$ are non-zero. We
will ignore other states that are either fully occupied or empty.
Furthermore, orthogonality of $\Phi_{1}$
and $\Phi_{2}$ leads to the simplification that only four vertices
contribute to the one- and two-particle propagators: $M_{\Phi_{1}\Phi_{1}\Phi_{1}\Phi_{1}}\left(0,q\right)$,
$M_{\Phi_{2}\Phi_{2}\Phi_{2}\Phi_{2}}\left(0,q\right)$, $M_{\Phi_{1}\Phi_{2}\Phi_{2}\Phi_{1}}\left(0,q\right)$,
and $M_{\Phi_{2}\Phi_{1}\Phi_{1}\Phi_{2}}\left(0,q\right)$.

The form of the vertices that are non-vanishing implies that $\mathcal{M}_{\Phi_{\alpha}\Phi_{\beta}}=\mathcal{M}_{\Phi_{\alpha}\Phi_{\alpha}}\delta_{\Phi_{\alpha}\Phi_{\beta}}$,
so from Eq.~(\ref{eq:SelfEnergy1}) the self-energy of the state
$\Phi$ is 
\begin{eqnarray*}
\Sigma_{\Phi} & = & -\frac{\nu_{\Phi}\mathcal{M}_{\Phi\Phi}}{1-\beta\nu_{\Phi}\left(1-\nu_{\Phi}\right)\mathcal{M}_{\Phi\Phi}}.\label{eq:SelfEnergyFirstOrder}\end{eqnarray*}
Since $\mathcal{M}$ is negative, the self-energy has no poles, and
hence is well-behaved for a contact potential.

We can also obtain simple expressions for the two-particle propagator.
The condition of orthogonality of $\Phi_{1}$ and $\Phi_{2}$ limits
the possible non-zero terms to $\Gamma_{\Phi_{1}\Phi_{1}\Phi_{1}\Phi_{1}}$,
$\Gamma_{\Phi_{1}\Phi_{2}\Phi_{2}\Phi_{1}}$, $\Gamma_{\Phi_{2}\Phi_{1}\Phi_{1}\Phi_{2}}$,
and $\Gamma_{\Phi_{2}\Phi_{2}\Phi_{2}\Phi_{2}}$. First, consider
the case of finite frequency. Taking the Fourier transform of each
expression with respect to $q$ with conjugate variable $s$, we can
solve for each of the two-particle propagators explicitly. We find
that $\tilde{\Gamma}_{\Phi_{1}\Phi_{1}\Phi_{1}\Phi_{1}}\left(s,\omega\right)=\tilde{M}_{\Phi_{1}\Phi_{1}\Phi_{1}\Phi_{1}}\left(0,s\right)$,
and $\tilde{\Gamma}_{\Phi_{2}\Phi_{2}\Phi_{2}\Phi_{2}}\left(s,\omega\right)=\tilde{M}_{\Phi_{2}\Phi_{2}\Phi_{2}\Phi_{2}}\left(0,s\right)$
which have no poles, and 
\begin{eqnarray}
\tilde{\Gamma}_{\Phi_{1}\Phi_{2}\Phi_{2}\Phi_{1}}\left(s,\omega\right) & = & \frac{\tilde{M}_{\Phi_{1}\Phi_{2}\Phi_{2}\Phi_{1}}\left(0,s\right)}{1+\tilde{M}_{\Phi_{1}\Phi_{2}\Phi_{2}\Phi_{1}}\left(0,s\right)\frac{\nu_{\Phi_{2}}-\nu_{\Phi_{1}}}{i\omega+\left(\xi_{\Phi_{2}}-\xi_{\Phi_{1}}\right)}},\mathbf{}\label{eq:Gamma1221}\end{eqnarray}
 \begin{eqnarray}
\tilde{\Gamma}_{\Phi_{2}\Phi_{1}\Phi_{1}\Phi_{2}}\left(s,\omega\right) & = & \frac{\tilde{M}_{\Phi_{2}\Phi_{1}\Phi_{1}\Phi_{2}}\left(0,s\right)}{1-\tilde{M}_{\Phi_{2}\Phi_{1}\Phi_{1}\Phi_{2}}\left(0,s\right)\frac{\nu_{\Phi_{2}}-\nu_{\Phi_{1}}}{i\omega-\left(\xi_{\Phi_{2}}-\xi_{\Phi_{1}}\right)}}.\label{eq:Gamma2112}\end{eqnarray}
The $\tilde{\Gamma}_{\Phi_{1}\Phi_{2}\Phi_{2}\Phi_{1}}\left(s,\omega\right)$
and $\tilde{\Gamma}_{\Phi_{2}\Phi_{1}\Phi_{1}\Phi_{2}}\left(s,\omega\right)$
propagators have poles at finite $\omega$ when $\nu_{\Phi_1} \ne \nu_{\Phi_2}$. Since
there are no poles in the $\mbox{Re}\left(\omega\right)>0$ half-plane,
we can analytically continue Matsubara frequencies to real frequencies,
$i\omega\to\omega+i\delta$. By symmetry,  $\tilde{M}_{\Phi_{1}\Phi_{2}\Phi_{2}\Phi_{1}}\left(0,s\right)=\tilde{M}_{\Phi_{2}\Phi_{1}\Phi_{1}\Phi_{2}}\left(0,s\right)$
so Eqs. (\ref{eq:Gamma1221}) and (\ref{eq:Gamma2112}) indicate
that the system is most susceptible to ordering in states $\Phi_1$ and $\Phi_2$ at the same 
wavelength $s_{{\rm CDW}}\left(\omega,T\right)$.

Let $\Delta\nu=\nu_{\Phi_{2}}-\nu_{\Phi_{1}}$ and $\Delta\xi=\xi_{\Phi_{2}}-\xi_{\Phi_{1}}$,
and assume $\xi_{\Phi_{2}}>\xi_{\Phi_{1}}$. After analytical continuation,
the frequency at which $\Gamma$ diverges is 
\begin{eqnarray*}
\omega_{c} & = & \Delta\xi\left[1+\tilde{M}_{\Phi_{1}\Phi_{2}\Phi_{2}\Phi_{1}}\left(0,s\right)\frac{\Delta\nu}{\Delta\xi}\right].\label{eq:OmegaCritical}
\end{eqnarray*}
At low temperatures, $\omega_{c}\propto$ $\beta$, which implies
that $\omega$ diverges as $T\to0$ unless $\omega\to0$
at finite temperature. This motivates us to consider the static ($\omega = 0$) limit.

In the static limit, all four two-particle vertices are modified by
the interaction. The vertices are
\begin{eqnarray}
\tilde{\Gamma}_{\Phi_{1}\Phi_{1}\Phi_{1}\Phi_{1}}\left(s,0\right) 
& = & \frac{\tilde{M}_{\Phi_{1}\Phi_{1}\Phi_{1}\Phi_{1}}\left(0,s\right)}{1-\beta\nu_{\Phi_{1}}\left(1-\nu_{\Phi_{1}}\right)\tilde{M}_{\Phi_{1}\Phi_{1}\Phi_{1}\Phi_{1}}\left(0,s\right)}, 
\label{eq:Gamma1111Static}
\end{eqnarray}
\begin{eqnarray}
\tilde{\Gamma}_{\Phi_{2}\Phi_{2}\Phi_{2}\Phi_{2}}\left(s,0\right) & = & \frac{\tilde{M}_{\Phi_{2}\Phi_{2}\Phi_{2}\Phi_{2}}\left(0,s\right)}{1-\beta\nu_{\Phi_{2}}\left(1-\nu_{\Phi_{2}}\right)\tilde{M}_{\Phi_{2}\Phi_{2}\Phi_{2}\Phi_{2}}\left(0,s\right)},
\label{eq:Gamma2222Static}
\end{eqnarray}
and
\begin{eqnarray}
\tilde{\Gamma}_{\Phi_{1}\Phi_{2}\Phi_{2}\Phi_{1}}\left(s,0\right) 
& = & \frac{\tilde{M}_{\Phi_{1}\Phi_{2}\Phi_{2}\Phi_{1}}\left(0,s\right)}{1+\frac{\nu_{\Phi_{2}}-\nu_{\Phi_{1}}}{\xi_{\Phi_{2}}-\xi_{\Phi_{1}}}\tilde{M}_{\Phi_{1}\Phi_{2}\Phi_{2}\Phi_{1}}\left(0,s\right)}, 
\label{eq:Gamma1221Static}
\end{eqnarray}
and by symmetry of the interaction vertices, $\tilde{\Gamma}_{\Phi_{1}\Phi_{2}\Phi_{2}\Phi_{1}}\left(s,0\right)=\tilde{\Gamma}_{\Phi_{2}\Phi_{1}\Phi_{1}\Phi_{2}}\left(s,0\right).$
We see in Eqs.~(\ref{eq:Gamma1111Static}) and (\ref{eq:Gamma2222Static})
that the system is unstable to CDW formation in $\Phi_{1}$ at a temperature
$$T_{11}=\nu_{\Phi_{1}}\left(1-\nu_{\Phi_{1}}\right)\tilde{M}_{\Phi_{1}\Phi_{1}\Phi_{1}\Phi_{1}}\left(0,s_1\right)$$
and in $\Phi_{2}$ at a temperature 
$$T_{22}=\nu_{\Phi_{2}}\left(1-\nu_{\Phi_{2}}\right)\tilde{M}_{\Phi_{2}\Phi_{2}\Phi_{2}\Phi_{2}}\left(0,s_2\right),$$ 
where $s_1$ and $s_2$ maximize $\tilde{M}_{\Phi_1\Phi_1\Phi_1\Phi_1}(0,s)$ and $\tilde{M}_{\Phi_2\Phi_2\Phi_2\Phi_2}(0,s)$  
respectively.

If $\xi_{\Phi_{2}} - \xi_{\Phi_{1}}$ is small, 
then the filling fractions $\nu_{\Phi_{1}}$ and $\nu_{\Phi_{2}}$
are almost equal. Expanding $\nu_{\Phi_{2}}$ about $\xi_{\Phi_{1}}$
to first order, $\nu_{\Phi_{2}}\approx\nu_{\Phi_{1}}+f'\left(\xi_{\Phi_{1}}-\mu\right)\left(\xi_{\Phi_{2}}-\xi_{\Phi_{1}}\right)$.
Then $\frac{\nu_{\Phi_{2}}-\nu_{\Phi_{1}}}{\xi_{\Phi_{2}}-\xi_{\Phi_{1}}}\approx-\beta\nu_{\Phi_{1}}\left(1-\nu_{\Phi_{1}}\right)+\mathcal{O}\left(f'''\left(\xi_{1}\right)\right).$
Eq.  (\ref{eq:Gamma1221Static}) allows us to find a third temperature scale, associated with 
interactions between states 1 and 2:
$$T_{12}=\nu_{\Phi_{1}}\left(1-\nu_{\Phi_{1}}\right)\tilde{M}_{\Phi_{1}\Phi_{2}\Phi_{2}\Phi_{1}}\left(0,s_{12}\right),$$
where $s_{12}$ maximizes $\tilde{M}_{\Phi_1\Phi_2\Phi_2\Phi_1}(0,s)$.
Whilst we have derived these results for
a contact potential, we will see that multiple temperature scales
also arise for a Coulomb potential in Sec.~\ref{sec:Numerics}.

\section{Landau theory for mixed CDW ordering}
\label{sec:Landau-free-energy-theory}
There has been considerable work on the Landau theory for interaction
induced CDWs in the lowest Landau level,\cite{FPA,Gerhardts} and
higher Landau levels.\cite{KFS,MoessnerChalker} Here we allow for
ordering of CDWs of states that arise from Landau level mixing and
focus on the situation in which there are two states close to the
Fermi energy.  We use similar notation to that of Fukuyama \emph{et
al.} (FPA).\cite{FPA}  Furthermore, we modify the notation of previous
sections and write $\sum_{\Phi_{i}}\to\sum_{i}$. 
The analysis in this section leads to a Landau theory where coupling between order
parameters of different states becomes important.

\subsection{Hartree-Fock Hamiltonian}
 
We consider an interaction $v(\bvec{q})$ and start from the interaction Hamiltonian 
 \begin{eqnarray}
H_{I} & = & \frac{1}{2}\sum_{\mathbf{q}\neq0}v\left(\bq\right)\left[
                  \hat{\rho}\left(\bq\right)\hat{\rho}\left(-\bq\right) 
                  + \hat{S}_x\left(\bq\right) \hat{S}_x\left(-\bq\right) 
                  - f\left(\bq\right)\right],
\end{eqnarray}
which is a natural generalization of the FPA interaction Hamiltonian to a situation with multiple states.\cite{Rasolt} The
operator $\hat{\rho}\left(\mathbf{q}\right)$ is the charge density operator of the two-state system, expressed as 
\begin{eqnarray}
\hat{\rho}\left(\mathbf{q}\right) & = & \sum_{i=1}^{2}\sum_{X}\rho_{iX}\left(\mathbf{q}\right)a_{iX_{+}}^{\dagger}a_{iX_{-}},
\end{eqnarray}
 where 
 \begin{eqnarray}
\rho_{iX}\left(\mathbf{q}\right) & = & \int d^2{\bvec{r}} \,  e^{-i\bvec{q}\cdot\bvec{r}} \phi^*_{iX}(\bvec{r}) \phi_{iX}(\bvec{r})
                                                  = e^{-iq_{x}X}A_{ii}(\mathbf{q}),
 \end{eqnarray}
where $\phi_{iX}(\bvec{r}) = \sum_\alpha C_{i\alpha} \psi_X^{(n_\alpha)}(\bvec{r})$, with $\psi$ defined in Eq.~(\ref{eq:LLwavefn}),
and with 
\begin{eqnarray}
A_{ij}\left(\mathbf{q}\right) & = & \sum_{\alpha}\sqrt{\frac{n_{ij_\alpha}'!} {\left( n_{ij_\alpha}'+\delta n_{ij_\alpha} \right)!}}C^*_{i\alpha}C_{j\alpha} e^{-i\left(n_{i_\alpha}-n_{j_\alpha}\right)\left(\theta-\pi/2\right)}
% \nonumber\\   &   & \times \, 
e^{-\frac{\left(ql_{0}\right)^{2}}{4}}\left[\frac{(ql_0)^2}{2}\right]^{\delta n_{ij}/2}\mathcal{L}^{\delta n_{ij_\alpha}}_{n_{ij_\alpha}'}\left[\frac{\left(ql_{0}\right)^{2}}{2}\right],
\end{eqnarray}
where $n_{ij_\alpha}' = \min\left(n_{i_\alpha}, n_{j_\alpha}\right)$, $\delta n_{ij_\alpha} = \left| n_{i_\alpha}-n_{j_\alpha} \right|$, and $\theta = \arctan\left(q_y/q_x\right)$. The operator $S_x\left(\mathbf{q}\right)$ is a pseudospin density operator
 \begin{eqnarray}
\hat{S}_x\left(\mathbf{q}\right) & = & \sum_{i,j=1}^{2}\sum_{X}s_{ij,X}\left(\mathbf{q}\right)a_{iX_{+}}^{\dagger}a_{jX_{-}},
 \end{eqnarray}
 where
 \begin{eqnarray}
s_{ij,X}\left(\mathbf{q}\right) & = & \sigma_{x}^{ij} \int d^2\bvec{r} \, e^{-i\bvec{q}\cdot\bvec{r}} \phi_{jX}^*(\bvec{r}) 
\phi_{iX}(\bvec{r}) =  \sigma_x^{ij}e^{-iq_{x}X}A_{ij}(\mathbf{q}),
 \end{eqnarray}
and $\sigma_x$ is the Pauli matrix. The first term in $H_I$ describes interactions between fermions 
in the same eigenstate, while the second term describes interactions between fermions in differing eigenstates.  Such a term can generically be expected
to be present in the situation we consider here,\cite{Rasolt} 
but usually will have a smaller magnitude than the first term 
in $H_I$ (which is the case here). The final term in $H_I$ is 
\begin{eqnarray}
f\left(\bq\right) & = & \sum_{i=1}^2 A_{ii}\left(\bq\right)^2 \rho_{i}(0).
\end{eqnarray}
We use different notation from the
notation in Sec.~\ref{sec:Moessner} by labeling states with $X=kl_{0}^{2}$
rather than pseudomomentum $k$, and $X_{\pm}=X\pm\frac{1}{2}q_{y}l_{0}^{2}$.
The indices $i$ and $j$ take values of 1 or 2 and label states $\Phi_{i}$
and $\Phi_{j}$. The operators $a^\dagger_{iX_\pm}$ ($a_{iX_\pm}$) create (annihilate) a fermion in the non-interacting eigenstate $\Phi_i$ with guiding centre coordinate $X_\pm = X \pm \frac{q_y l_0^2}{2}$, and satisfy the usual anticommutation relations $\left\{a_{iX_\pm},a^\dagger_{jY_\pm}\right\} = \delta_{ij}\delta_{X_\pm,Y_\pm}$. In the limit of no mixing, and when restricted to
the lowest LL, the interaction Hamiltonian reduces to the form considered
by FPA. 

The total number of fermions in the two-state system is $N = N_1 + N_2 = \frac{L^2}{2\pi l_0^2}$, where 
\begin{eqnarray}
N_i & = & \left<\sum_X a_{iX}^\dagger a_{iX}\right>,
\end{eqnarray}
is the number of electrons in state $\Phi_i$. Similarly to FPA, we define the order parameter for a CDW in eigenstate
$\Phi_{i}$, $\Delta_{i}\left(\mathbf{Q}_{i}\right)$ by
\begin{eqnarray}
\left\langle a_{iX_{+}}^{\dagger}a_{jX_{-}}\right\rangle  & = & \frac{2\pi}{L}\delta_{ij}P_i\sum_{\mathbf{Q}_{i}}\left[\Delta_{i}^{*}\left(\mathbf{Q}_{i}\right)e^{i\left(Q_{i}\right)_{x}X}\delta\left(\left(Q_{i}\right)_{y}-q_{y}\right) 
%\right. \nonumber\\ & & \left. 
+ \Delta_{i}\left(\mathbf{Q}_{i}\right)e^{-i\left(Q_{i}\right)_{x}X}\delta\left(\left(Q_{i}\right)_{y}+q_{y}\right)\right],
\label{eq:OrderParameter}
\end{eqnarray}
where $P_i = N_i/N$ is the fraction of the fermions that occupy state $\Phi_i$. Each order parameter depends on a unique set of wavevectors
$\left\{ \mathbf{Q}_{i}\right\} $, where $\mathbf{Q}_{i}=\left(\left(Q_{i}\right)_{x},\left(Q_{i}\right)_{y}\right)$.
The summation over wavevectors allows for CDW ordering
at several wavevectors. For instance, a triangular
CDW can be described by a set of three wavevectors of equal length,
oriented at angles $\frac{2\pi}{3}$ with respect to each other,\cite{FPA}
whereas a striped CDW is described by a single wavevector.\cite{MoessnerChalker}
Furthermore, note that the order parameters are defined such that
$\Delta_{i}^{*}\left(\mathbf{Q}_{i}\right)=\Delta_{i}\left(-\mathbf{Q}_{i}\right)$,
which follows from Eq.~(\ref{eq:OrderParameter}).  

Next, we use the Hartree-Fock approximation to rewrite the Hamiltonian
as: 
\begin{eqnarray}
H & = & H_0 +  H_{{\rm CDW}}+N\sum_{i=1}^{2}\sum_{\left\{ \mathbf{Q}_{i}\right\} }P_i^2 U_{ii}\left(Q_{i}\right)|\Delta(Q_{i})|^{2}, \nonumber \\ & & 
\end{eqnarray}
where $$H_0 = \sum_{i=1}^2\sum_X \epsilon_i^0 a_{iX}^\dagger a_{iX},$$ is the non-interacting Hamiltonian with $\epsilon_{i}^{0}$ 
the energy of single particle eigenstate $\Phi_{i}$, and

 \begin{eqnarray}
H_{\rm CDW} 
& = & -\left\{\sum_{i=1}^{2}\sum_{\left[ \mathbf{Q}_{i}\right\} }P_iU_{ii}\left(Q_{i}\right)\sum_{X}\left[ \Delta_{i}\left(\mathbf{\mathbf{Q}_{i}}\right)e^{i(Q_i)_{x}X}a_{iX_{-}}^{\dagger}a_{iX_{+}}+\Delta_{i}^{*}\left(\mathbf{\mathbf{Q}_{i}}\right)e^{-i(Q_i)_{x}X}a_{iX_{+}}^{\dagger}a_{iX_{-}}\right] \right.\nonumber \\
&    & \left.+\sum_{i=1}^{2}\sum_{j\neq i}P_{i}U_{ij}\left(Q_{i}\right)\sum_{X}\left[ \Delta_{i}\left(\mathbf{Q}_{i}\right)e^{i(Q_i)_{x}X}a_{jX_{-}}^{\dagger}a_{jX_{+}}+\Delta_{i}^{*}\left(\mathbf{Q}_{i}\right)e^{-i(Q_i)_{x}X}a_{jX_{+}}^{\dagger}a_{jX_{-}}\right] \right\}.\nonumber \\ 
&    &{\rm }
\label{eq:HamiltonianHartreeFock}
\end{eqnarray}  
We let $Q_{i}=\left|\mathbf{Q}_{i}\right|$
and $\theta_{i}=\arctan\left[\frac{\left(Q_{i}\right)_{y}}{\left(Q_{i}\right)_{x}}\right]$
and henceforth we absorb $l_0$ in the definition of $q$, i.e. $ql_0 \to q$. 
In the summations over wavevectors, we sum over $\left\{ \mathbf{Q}_{i}^{\alpha}\right\} _{\alpha=1}^{n}$,
where $n$ is the number of wavevectors describing the CDW state.
Furthermore, we define the following functions, which describe the
Hartree-Fock interaction potentials: 
 \begin{eqnarray*}
U_{ii}\left(Q_{i}\right) & = & \int\mbox{d}q\hspace{0.05cm} q\hspace{0.05cm}  v\left(q\right)A_{ii}\left(q\right)^{2}J_{0}\left(qQ_{i}\right) -v\left(Q_{i}\right)A_{i}\left(Q_{i}\right)^{2},\\
U_{ij}\left(Q_{i}\right) & = &  \frac{1}{2\pi}\sum_{m=-\infty}^{\infty}\int\mbox{d}^2 \mathbf{q}\hspace{0.05cm}  v\left(q\right)A_{ij}\left(q\right)A_{ji}\left(q\right)J_{m}\left(qQ_{i}\right)-v\left(Q_{i}\right)A_{i}\left(Q_{i}\right)A_{j}\left(Q_{i}\right),
\end{eqnarray*}
The potentials $U_{ii}$ and $U_{ij}$ represent 
the most significant contributions to the Hartree-Fock Hamiltonian. 
We ignore other terms generated by the Hartree-Fock approximation, 
as their contributions are negligible.

\subsection{Landau theory}

We construct a Landau theory in the spirit of FPA by considering the difference in free
energy between a uniform system and an ordered system for a
fixed total number of electrons, $N$. We write the difference in free energy,
$\delta F$, in terms of the grand potentials of each system, 
\begin{equation*}
\delta F=\Omega_{\rm CDW}\left(\mu\right)-\Omega_{0}\left(\mu\right)+\frac{1}{2}\left(\mu-\mu_{0}\right)^{2}\frac{\partial^{2}\Omega_{0}\left(\mu_{0}\right)}{\partial\mu_{0}^{2}},
\end{equation*}
 where $\Omega_{\rm CDW}$ and $\Omega_{0}$ are the grand potentials
of the CDW and uniform systems, respectively, and the chemical potentials
of the uniform and CDW systems are $\mu_{0}$ and $\mu$, respectively,
and we expand in powers of the CDW order parameters $\Delta_1\left(\mathbf{Q}_1\right)$ and $\Delta_2\left(\mathbf{Q}_2\right)$.

The grand potential when there is CDW ordering is 
\begin{eqnarray*}
\Omega_{\rm CDW}\left(\mu\right) & = & -\beta^{-1}\ln\left\{\mbox{Tr}\left[e^{-\beta\left(H -\mu\sum_{i}\sum_{X}\hat{n}_{iX}\right)}\right]\right\}\nonumber \\
 & = & \Omega_{0}\left(\mu\right)-\beta^{-1}\ln\left\langle e^{-\beta H_{\rm CDW}}\right\rangle _{0}+N\sum_{i=1}^{2}\sum_{\left\{ \mathbf{Q}_{i}\right\} }P_i^2U_{ii}\left(Q_{i}\right)\left|\Delta_{i}\left(\mathbf{Q}_{i}\right)\right|^{2},\label{eq:ThermoPotentialCDW}
 \end{eqnarray*}
 where $\left<\ldots\right>_{0}$ indicates an average with respect
to the uniform state, $H_{{\rm CDW}}$ is defined above in Eq.~(\ref{eq:HamiltonianHartreeFock}),
 and 
 \begin{eqnarray*}
\Omega_{0}\left(\mu\right) & = & -N\beta^{-1}\sum_{i}\ln\left[1+e^{-\beta\left(\epsilon_{i}^{0}-\mu\right)}\right].
\end{eqnarray*}

We now develop the perturbative expansion of the free energy in terms of the order
parameters $\Delta_{1}\left(\mathbf{Q}_{1}\right)$ and $\Delta_{2}\left(\mathbf{Q}_{2}\right)$. We expand the
logarithm, which gives 
\begin{eqnarray}
\frac{\delta F}{N} 
& = & \sum_{i=1}^{2}\sum_{\left\{ \mathbf{Q}_{i}\right\} }P_i^2U_{ii}\left(Q_{i}\right)\left|\Delta_{i}\left(\mathbf{Q}_{i}\right)\right|^{2}
-\frac{1}{N}\sum_{p=1}^{\infty}\frac{\left(-1\right)^{p}\beta^{p-1}}{p!}\left\langle \left(H_{\rm HF}\right)^{p}\right\rangle _{0}+\frac{1}{2N}\left(\mu-\mu_{0}\right)^{2}\frac{\partial^{2}\Omega_{0}\left(\mu_{0}\right)}{\partial\mu_{0}^{2}}.\label{eq:ThermoPotentialCDWExpand}\end{eqnarray}
 The $p=1$ term in the expansion vanishes, since it involves averages
of the form $\left\langle a_{iX_{+}}^{\dagger}a_{iX_{-}}\right\rangle _{0}=\delta_{X_{+}X_{-}}$
which imply $\mathbf{Q}_{i}=0$. We consider Eq.~(\ref{eq:ThermoPotentialCDWExpand}) to fourth order
in both order parameters and write the CDW order parameters in the
form $\Delta_{k}=\Delta_{k}\left(\mathbf{Q}_{k}\right)$.
The general form of the free energy density when $\bvec{Q}_{1}\neq\bvec{Q}_{2}$
is \begin{eqnarray}
\mathcal{F} & = &  P_1^2 a_{1}\left(\mathbf{Q}_{1},T\right)|\Delta_{1}|^{2} 
                           + P_2^2 a_{2}\left(\mathbf{Q}_{2},T\right)|\Delta_{2}|^{2}
                           + P_1^3 b_{1}\left(\mathbf{Q}_{1},T\right)|\Delta_{1}|^{3} 
                           + P_2^3 b_{2}\left(\mathbf{Q}_{1},T\right)|\Delta_{2}|^{3} \nonumber \\ 
                   &  &  \hspace{0.1cm} + P_1^4 c_{1}\left(\mathbf{Q}_{1},T\right)|\Delta_{1}|^{4} 
                           + P_2^4 c_{2}\left(\mathbf{Q}_{2},T\right)|\Delta_{2}|^{4}
                           + P_1^2 P_2^2\gamma\left(\mathbf{Q}_{1},\mathbf{Q}_{2},\theta,T\right)|\Delta_{1}|^{2}|\Delta_{2}|^{2},
\label{eq:FourthOrderFreeEnergy}
\end{eqnarray}
where $\mathcal{F}=\delta F/N$. We calculate the
coefficients $a_{i}$, $b_{i}$, $c_{i}$ and $\gamma$ explicitly
within the Hartree-Fock approximation allowing for the possibility
of either striped or triangular CDW order in either state 1 or state
2. The striped phase is described by a single wavevector while the triangular
CDW is described by a set of three wavevectors of equal magnitude,
oriented at 120$^{\circ}$ with respect to each other.
 The expressions for these coefficients are quite lengthy and we
present them in full detail in Appendix~\ref{sec:Landau-free-energy-coefficients}.
The phase behaviour of free energies of the form in 
Eq.~(\ref{eq:FourthOrderFreeEnergy}) when $b_{1}=b_{2}=0$ was studied
in detail by Imry.\cite{Imry} In general, there are three possible
critical points, with free energies $\mathcal{F}_{1}$, $\mathcal{F}_{2}$,
and $\mathcal{F}_{M}$. If $\mathcal{F}_{1}$ is globally minimum,
then $\Delta_{1}\neq0$, and $\Delta_{2}=0$ and vice versa for $\mathcal{F}_{2}$.
If $\mathcal{F}_{M}$ is the global minimum, then both $\Delta_{1}\neq0$
and $\Delta_{2}\neq0$. In the case we consider, the phases associated
with $\mathcal{F}_{1}$ and $\mathcal{F}_{2}$ are CDW ground states
in either state $\Phi_{1}$ or $\Phi_{2}$, respectively. The phases
associated with $\mathcal{F}_{M}$ are mixed CDW states. The mixed CDW
phase exists only when the parameter $\gamma$, which couples ordering
in states 1 and 2 is finite. If $\gamma>0$, then there is competition
between order parameters of each state; if $\gamma<0$, then ordering
in one state serves to enhance ordering in the other.\cite{Imry}
 The phases of the order parameters are important
only for triangular CDWs.\cite{FPA} If the phases corresponding to
the three different order parameters are denoted by
$\left\{ \phi_{i}\right\} _{i=1}^{3}$, then for a state $\Phi$,
$\phi_{1}+\phi_{2}+\phi_{3}=0\mbox{ }(\pi)$ if $\nu_{\Phi}<1/2$ $(>1/2)$.

There are additional terms in the free energy that are allowed when
$\bvec{Q}_{1}=\bvec{Q}_{2}$, which take the form 
\begin{eqnarray}
\delta{\mathcal{F}}_{\bvec{Q}_{1}=\bvec{Q}_{2}} & = & P_1 P_2\alpha\left|\Delta_1\right|\left|\Delta_2\right| 
                                                                                     + P_1^2 P_2\eta\left|\Delta_1\right|^2\left|\Delta_2\right| 
                                                                                     + P_1 P_2^2\varphi\left|\Delta_1\right|\left|\Delta_2\right|^2 
                                                                                     + P_1^2 P_2^2\tilde{\gamma}\left|\Delta_1\right|^2\left|\Delta_2\right|^2 \nonumber \\ 
                                                                             &    & \hspace{0.1cm} + P_1^3 P_2\sigma\left|\Delta_1\right|^3\left|\Delta_2\right| 
                                                                                     + P_1 P_2^3 \rho\left|\Delta_1\right|\left|\Delta_2\right|^3.
\label{eq:deltafree}
\end{eqnarray}
The coefficients $\alpha$, $\eta$,
$\varphi$, $\tilde{\gamma}$, $\sigma$, and $\rho$
 depend on the magnitude of the ordering wavevectors $Q_{1}=Q_2 = Q$ and the relative phases of the order parameters. Note that $\tilde{\gamma}$ modifies $\gamma\left(\mathbf{Q},\mathbf{Q},0,T\right)$ in the case of equal wavevectors. These coefficients  are displayed
in Appendix~\ref{sec:Landau-free-energy-coefficients-q1=00003D=00003Dq2}.

We adopt the following notation for describing the possible phases.
 For ordering in a single state, we label according to the symmetry and 
state index. For example, striped (triangular) ordering in $\Phi_{1}$ is denoted
S1 (T1). For mixed CDW ordering, we label according to the symmetry
of the order parameter for each phase with order in $\Phi_1$ preceding 
that in $\Phi_2$. For example, mixed CDW ordering
between striped phases in both states is labeled SS. If the wavevectors
for each phase are equal, we add an extra ``E'' to the end of
the label. In all, we allow for the possibility of 12 types of CDW
ordering: S1, S2, T1, T2, SS, TT, ST, TS, SSE, TTE, STE, and TSE.
We present results for specific numerical examples in Sec.~\ref{sec:Numerics}.

\section{Numerics}

\label{sec:Numerics}

The Landau theory developed in Sec.~\ref{sec:Landau-free-energy-theory}
makes no assumptions about the source of LL mixing, and does not
specify the interaction $v(\bvec{q})$.  There are many 
possible sources of LL mixing in 2DEGs that can affect CDW order,
such as electron-electron interactions\cite{MacDonald} or
 disorder.\cite{Stanescu}
However, we focus on the specific example of Rashba spin-orbit coupling, since 
this may give a way to tune 
levels into degeneracy controllably and create a situation of the form 
discussed in Sec.~\ref{sec:Landau-free-energy-theory}.  Electron-electron
interactions are described by an unscreened Coulomb potential -- 
we expect that screening may lead to small quantitative changes in 
our results, but that our qualitative results will be robust as
we discuss in Sec.~\ref{sec:Screening}

\subsection{Model and parameters}

\subsubsection{Rashba spin-orbit coupling}

\label{sec:Rashba}

 Rashba spin orbit coupling can be significant in GaAs
quantum wells and leads to  LL mixing.
For holes near the $\Gamma$ point in GaAs there are four $j=3/2$ spin
states in the valence band, which separate into two heavy-hole
(HH) and two light-hole (LH) states with different effective masses. The
heavy hole states correspond to angular momentum $j_{z}=\pm3/2$
and the light hole states correspond to $j_{z}=\pm1/2$. 
The strength of Rashba coupling
can be tuned through external means, such as a back gate,
which allows tuning of levels into and out of degeneracy. Generically,
Dresselhaus spin-orbit coupling will also be present in GaAs, however,
we focus on the regime where Rashba coupling is dominant and ignore
additional (non-tunable) mixing that may arise from Dresselhaus
coupling.

We specify the single-particle Hamiltonian, which includes Rashba coupling
and Zeeman splitting, as \begin{eqnarray}
H_{\rm non-interacting} & = & H_{0}+H_{z}+H_{SO},\label{eq:HamiltonianTotal}\end{eqnarray}
 and diagonalize to find the non-interacting eigenstates as a function
of the strength of the magnetic field and the Rashba coupling. We
then consider a specific example in which there are two energy levels
that are close to degeneracy at a particular value of Rashba coupling, and explore
the effects of interactions on the phase diagram as a function of
temperature, filling and Rashba coupling.

We take the free-particle Hamiltonian to be
\begin{eqnarray*}
H_0 = \sum_{j_z} \hbar \omega(|j_z|) \left(\hat{n}_{j_z} + \frac{1}{2}\right),
\label{eq:HamiltonianFree}
\end{eqnarray*}
where $$ \omega(|j_z|) = \left\{\begin{array}{ccc} \omega_{HH} & {\rm for} & j_z =\pm \frac{3}{2}
\\ \\ \omega_{LH} & {\rm for} & j_z = \pm\frac{1}{2} \end{array} \right. ,$$
and the number operator $\hat{n}_{j_z} = a^\dagger_{j_z}a_{j_z}$, with
$a^\dagger_{j_z}$ the LL raising operator for a hole with $J_z$ eigenvalue $j_z$.
 $\omega_{HH}$ and $\omega_{LH}$ are the cyclotron frequencies of the 
heavy and light hole states, respectively. 
The Zeeman splitting term is given by 
\begin{eqnarray*}
H_{Z} & = & -g^* \mu_B B \hat{J}_z,
\end{eqnarray*}
where the effective $g$-factor for the holes is $g^*$. 
 The third operator may be written as
\begin{eqnarray*}
H_{SO} & = &  -\alpha \mathcal{R}_{8v8v}\left\langle \mathcal{E}_{z}^{0}\right\rangle
\left(\bvec{k}\times\bvec{J}\right)_z ,
\label{eq:HamiltonianSO}
\end{eqnarray*}
which introduces Rashba spin-orbit coupling into the Hamiltonian.
The parameter $\mathcal{R}_{8v8v}$ is a material-dependent Rashba coefficient;
$\left\langle \mathcal{E}_{z}^{0}\right\rangle =\frac{e}{2\epsilon}\rho$
is the electric field due to the fermions in the absence of a 
substrate field.\cite{WinklerBook}
We use parameters that are closely related to those relevant to
Fischer {\it et al.}'s experiments.
We choose material parameters $g^* = 1.2$, 
$\mathcal{R}_{8v8v} = 14.62 \,  e {\rm \AA}^2$, and $\epsilon = 12.4 \, \epsilon_0$,
appropriate to bulk GaAs;\cite{WinklerBook} we set the
heavy- and light-hole masses equal to $m_{H}=0.51m_{0}$ and $m_{L}=0.08m_{0}$,
respectively;\cite{GaAsParameters} and we set the in-plane carrier density to
 $\rho=3.5\times10^{11}$ cm$^{-2}$.\cite{Schmult}
 
\begin{figure}
\includegraphics[width=3in]{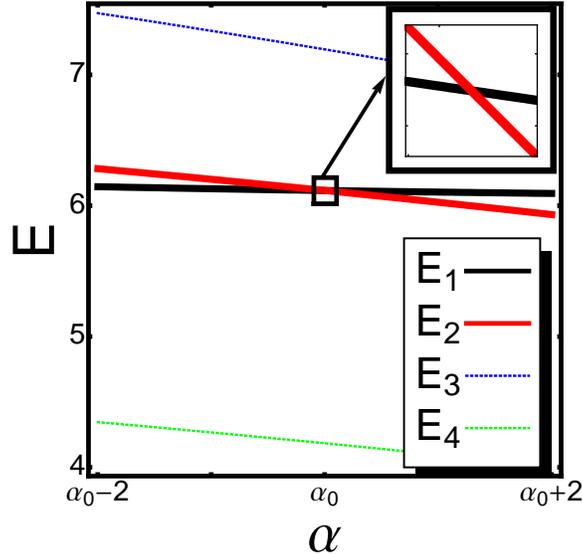}
\caption{\label{fig:EnergyVersusAlpha}Plot of energy of non-interacting eigenstates
$\Phi_{1}$ and $\Phi_{2}$ versus tuning parameter $\alpha$ at $\nu_{T}=7.5$.
Energy is in units of $E_{0}=\hbar eB/m_{0}$ and $\alpha$ is dimensionless.
The nearest energy levels  above and below $E_{1}$ and $E_{2}$ are labeled
$E_4$ and $E_3$, respectively. Inset: Energy levels $E_1$ and $E_2$ for
$\alpha \in [\alpha_0 - 0.01,\alpha_0 + 0.01]$.}
\end{figure}

We model the effect of tuning via external means 
(such as a back gate) by the parameter
$\alpha$. The magnetic field $B$ is fixed according
to the total filling fraction of the system, $\nu_{T}=2\pi\rho l_{0}^{2}$.
The term $\alpha\left\langle \mathcal{E}_{z}^{0}\right\rangle $ may
be compared to the average confinement electric field seen by fermions
in a triangular potential, as discussed by Fischer {\it et al}.\cite{Grayson}
The parameter $\alpha$ then corresponds to the enhancement of the
average electric field by increasing the substrate electric field,
$\mathcal{E}_{s}$. That is, $\alpha  =  1+\frac{\mathcal{E}_{s}}{\left\langle \mathcal{E}_{z}^{0}\right\rangle }.$
 For no substrate field, $\alpha=1$, whereas in the Fischer {\it et al.} experiments,
$\alpha\approx$ 2-3 corresponding to nonzero $\mathcal{E}_{s}$.
For the levels we study here, a crossing takes place for 
$\alpha_0 \simeq 20$.  The values of $\alpha$ at which level crossings
occur are very sensitive to the model parameters.  In particular, the
values of the material parameters, such as the effective masses and the carrier density, and the presence or absence of
Dresselhaus coupling can alter the positions of crossings and the 
nature of the levels that cross dramatically.  Fischer {\it et al}.
construct a Hamiltonian from the $\Gamma_6^c$, $\Gamma_8^v$, and $\Gamma_7^v$ 
bands, whereas we consider the $\Gamma_8^v$ band only. Second, 
they include both the Rashba and Dresselhaus spin-orbit coupling terms, 
whereas we consider Rashba coupling only.  Third, we set the carrier density to the upper 
limit studied by Schmult \emph{et al.},\cite{Schmult} which is larger than any 
of the densities studied by Grayson \emph{et al.}  This choice of carrier density 
increases the energy gap between the energy levels comprising the two-state system 
and the nearby energy levels, as shown in Fig. (\ref{fig:EnergyVersusAlpha}), 
thereby allowing us to focus on states $\Phi_1$ and $\Phi_2$ only.  Whilst our numerical
calculations are not adequate for a quantitative discussion of the
experiments in Ref.~\onlinecite{Grayson}, the crossings we are 
able to induce with increasing $\alpha$ in our simplified model 
should lead to 
a qualitative understanding of the effect of LL
mixing on CDW formation.  As we point out in Sec.~\ref{sec:Genfeatures}
there are certain features in the phase diagram we expect to be 
generic and others that may depend more sensitively on details of LL mixing.

We diagonalize the Hamiltonian shown in Eq.~(\ref{eq:HamiltonianTotal}) 
and find the eigenvectors and associated eigenvalues as functions of $\alpha$
and magnetic field.  The non-linear dependence of the energy levels on 
magnetic field allows for energy crossings.\cite{WinklerBook} 
Cubic and tetrahedral corrections can modify these crossings to
anti-crossings.\cite{Grayson} 
Motivated by the results of Fischer \emph{et al.}\cite{Grayson} 
we focus on 
parameter values at which there are two energies that are close 
to degenerate and located near the Fermi energy.  
This is the situation we discussed in Sec.~\ref{sec:Landau-free-energy-theory}.
In the vicinity of such a crossing, we 
write the filling fraction $\nu=\bar{\nu}+\nu_{0}$, where $\bar{\nu}$ is an integer, corresponding to the number of
fully occupied energy levels, and $\nu_{0} = \nu_1 + \nu_2$ is the 
total filling fraction of the two partially filled energy levels.
 We label the crossing states $\Phi_{1}$ and $\Phi_{2}$,
their associated energies $E_{1}$ and $E_{2}$, and their filling
fractions $\nu_{1}$ and $\nu_{2}$. 
Our analysis is not
 greatly affected if crossings are replaced by anti-crossings, as the
key point is that there be non-zero filling in the two states, which will
be the case when the temperature is not too low for an anti-crossing.
As $T\to0$, only the lowest energy state is occupied, which
decouples the two states, except when $E_1 = E_2$.

\subsubsection{Interactions}
\label{sec:Screening}
We choose the interactions between holes to be given by
the unscreened Coulomb interaction, similarly to FPA\cite{FPA} \[
v\left(\mathbf{q}\right)=\frac{2\pi}{A}\frac{e^{2}}{4\pi\epsilon\left|\mathbf{q}\right|},\]
 where $A$ is the area of the 2DHS (the 
factor $2\pi/A$ comes from the Fourier transform of the real-space Coulomb potential). 
This allows for a somewhat more realistic interaction than the contact
potential considered in Sec.~\ref{sec:Moessner}. 

When there are multiple filled Landau levels, the filled levels can act as a 
background that screens the bare Coulomb interaction.  
It was found by Aleiner and Glazman\cite{AleinerGlazman} (AG)
that the screening from filled Landau levels can be accounted for by replacing $\epsilon$ with 
\begin{eqnarray}
\epsilon_{AG}\left(q\right) & = & \left\{1+\frac{2}{q a_{B}}\left[1-J_{0}\left(qR_{L}\right)^{2}\right]\right\}\epsilon_0,
\label{eq:AGDielectric}
\end{eqnarray}
where $R_{L} = v_F/\omega_c$ and $a_{B}$ are the Larmor radius and Bohr radius,
respectively. Fogler, Koulakov and Shklovskii\cite{FKS} used 
this screened interaction  
to show that electrons in a partially filled LL near half-filling preferentially order in a striped CDW phase in a weak magnetic field. Away from half-filling, they found  ordering in a triangular CDW phase.\cite{FKS} 

In the $n^{\rm th}$ LL at wavevectors $ql_{0}\approx1/\sqrt{n}$, $\epsilon_{AG}\approx R_{L}/a_{B}$;
at low magnetic fields, $\epsilon_{AG}$ is large and screening effects
are appreciable. However, at wavevectors $ql_{0}\gtrsim1$ or $ql_{0}\lesssim1/n$,
screening is negligible. In the example considered in this paper,
the CDW ordering generally occurs when $ql_{0}\gtrsim 1$
for an unscreened Coulomb interaction.
This is outside the region where the effects of screening would be
seen clearly if we considered the modified dielectric constant in
Eq.~(\ref{eq:AGDielectric}). For the purpose of this work, we consider
the bare Coulomb interaction as we focus on non-interacting states 
for which the dominant constituent LLs have $n \leq 4$, hence
we expect our phase diagrams to be qualitatively 
correct, although screening may lead to small quantitative corrections.

\subsubsection{States}
\label{sec:states}
Our approach to finding phase diagrams numerically starts by
identifying a pair of states which have a crossing.  
In Fig.~\ref{fig:EnergyVersusAlpha}, we plot the energy
of states that we label $\Phi_{1}$ and $\Phi_{2}$ for a magnetic field $B$ corresponding to $\nu_{T}=8$ for our chosen electron density.  In addition to these
states, we also plot the nearest energy levels above and below $\Phi_{1}$
and $\Phi_{2}$. At $B\approx1.8$ T 
the minimum energy gap to the  levels above and below
$\Phi_{1}$ and $\Phi_{2}$ is approximately 0.23 meV $\sim 2.7$ K, which is
 large compared to the spacing between $\Phi_1$ and $\Phi_2$, which is 
approximately 2.3 $\mu$eV.   In comparison, the transition temperature we calculate for
the CDW phase is $\sim 10$ K.  The Hartree-Fock approximation thus overestimates the
transition temperatures observed in experiment, in which 
anisotropic transport develops for $T \lesssim 100$ mK.\cite{Dorsey1}  Nevertheless,
we ignore adjacent energy levels and construct a two-state system from the crossing energy levels, and make only qualitative predictions about the possibility of CDW ordering as a function of $\nu_T$ and $\alpha$. In the situation shown in Fig.~\ref{fig:EnergyVersusAlpha} there
are seven filled levels, and we investigate CDW ordering in the two
states near the Fermi energy for $0.5\le\nu_0\le 1$.

\subsection{Quantum Hall Ferromagnetism}
\label{sec:QHF}
We have focused primarily on the possibility of CDW ordering when 
two energy levels are in the vicinity of the Fermi energy, and each level has 
nonzero filling. When two levels are degenerate or close to degenerate then the
physics of quantum Hall pseudospin ferromagnetism can also be 
important.\cite{MPB,JungwirthMacdonald}  
Following Jungwirth and Macdonald,\cite{JungwirthMacdonald} we determine
the Landau theory for quantum Hall pseudospin ferromagnetism when there is 
Landau level mixing using the Hartree-Fock approximation.  
This allows us to compare the ordering
temperature for quantum Hall ferromagnetism with the temperature we have 
determined for CDW formation in the vicinity of degeneracy for the 
non-interacting states. 

We define the psuedospin operators in terms of the 
creation and annihilation operators of the non-interacting eigenstates: 
\begin{eqnarray}
\label{eq:QHFoperators}
\hat{m}_{I,\mathbf{q}} & = & \sum_{X}a^\dagger_{iX_+} \sigma_I^{(ij)} a_{jX_{-}}e^{-iq_xX},
\end{eqnarray}
where $\sigma_I$ is the Pauli spin matrix and $I=0$, $x$, $y$, and $z$ are pseudospin labels ($\sigma_0$ is the two-dimensional identity operator). By construction, $\hat{m}^\dagger_{I,q_y} = \hat{m}_{I,-q_y}$. We focus on $\bvec{q}=0$ order, so 
the order parameter is defined as
\begin{eqnarray}
\label{eq:QHFOrderParameters}
m_I & = & \frac{2\pi}{L}\delta_{\mathbf{q},0}\left<a^\dagger_{iX_+} \sigma_I^{(ij)} a_{jX_{-}}\right>.
\end{eqnarray}
Since the total filling fraction of the two-state system is fixed, $m_0 = \nu_T$.

We rewrite the creation and annihilation operators of states $\Phi_1$  and $\Phi_2$ in terms of the pseudospin operators and proceed using the Hartree-Fock approximation in the same manner as outlined in Sec.~\ref{sec:Landau-free-energy-theory} or 
Ref.~\onlinecite{JungwirthMacdonald}. 
This leads to the effective Hamiltonian 
\begin{eqnarray}
\label{eq:QHFHamiltonian}
H_{QHF} & = & \sum_{I=0,z}b_I m_I - \sum_{IJ=0,x,y,z}W_{IJ} m_I \hat{m}_{J,0}  + \frac{N}{2}\sum_{IJ=0,x,y,z}W_{IJ} m_I m_J,
\end{eqnarray}
where $b_0 = \frac{\epsilon_1^0+\epsilon_2^0}{2} - \frac{1}{2}\sum_{I}W_{II}$ and 
$b_z = \frac{\epsilon_1^0-\epsilon_2^0}{2} - W_{0z}$. 
The potentials $W_{IJ}$ are defined explicitly in Appendix \ref{sec:HFQHF}. 
We note that for the eigenstates relevant to the numerical example shown below,  
$W_{0I} = W_{zI} = 0$, $I = x,$ $y$, and so $m_0$ and $m_z$ 
do not couple to $m_x$ and $m_y$.

To second order in the pseudospin order parameters, the change in 
free energy associated with ordering in a QHF takes the form:
\begin{eqnarray}
\delta f_{QHF} &=&\sum_{I=0,z}b_I m_I  + \sum_{IJ = 0,z}W_{IJ}m_I\left<\hat{m}_{J,0}\right>_0\nonumber\\
                        & & + \frac{1}{2}\sum_{IJ}W_{IJ} m_I m_J-\frac{\beta}{2}\sum_{IK,JL}W_{IK}W_{JL}m_{I}m_{J}\left<\hat{m}_{K,0}\hat{m}_{L,0}\right>_0.
\end{eqnarray}
The $\left<...\right>_0$ notation indicates that we average over non-interacting 
states. Thus, $\left<\hat{m}_{0,0}\right>_0 = \nu_1 + \nu_2 =\nu_T$ 
and $\left<\hat{m}_{z,0}\right>_0 = \nu_1 - \nu_2 = \delta\nu$.  It follows that 
\begin{eqnarray}
\delta f_{QHF} &=&\frac{1}{2}W_{00} m_0^2 \left[1 + \left(f'_1+f'_2\right)\left(W_{00}+\frac{W_{z0}^2}{W_{00}}\right)+ \left(f'_1-f'_2\right)W_{0z}\right]\nonumber\\
                        &  &+ \frac{1}{2}W_{xx} m_x^2 \left[1 - \beta \left(\nu_T - 2\nu_1\nu_2\right)W_{xx}\right]\nonumber\\
                        &  &+ \frac{1}{2}W_{yy} m_y^2 \left[1 - \beta \left(\nu_T - 2\nu_1\nu_2\right)W_{yy}\right]\nonumber\\
                        &  &+ \frac{1}{2}W_{zz} m_z^2 \left[1 + \left(f'_1+f'_2\right)\left(W_{zz}+\frac{W_{z0}^2}{W_{zz}}\right)+ \left(f'_1-f'_2\right)W_{0z}\right]\nonumber\\
                        &  &+\left[\nu_T W_{00}+\delta\nu W_{0z} - b_0\right]m_0 + \left[\nu_T W_{0z} + \delta\nu W_{zz} + b_z\right]m_z \nonumber\\
                        &  &+\left[W_{0z}\left(f'_1+f'_2\right)\left(W_{00}+U_{zz}\right)+\left(f'_1-f'_2\right)\left(W_{00}U_{zz}+W_{0z}^2\right)\right]m_0 m_z.
\end{eqnarray}
Since $m_0$ is constant, $\partial\delta f_{QHF}/\partial m_0 = 0$. Thus, terms containing $m_0$ only act to shift the minimum of the free energy, and can be ignored safely. Let us then write the remaining terms as follows:
\begin{eqnarray}
\delta f_{QHF} &=& \frac{1}{2}\sum_{I=x,y,z}a_{I}\left(T\right)W_{II} m_I^2 + \tilde{b}_z\left(T\right)m_z.\nonumber
\end{eqnarray}
The coefficients $a_{I}$  determine the 
temperature scales for a second-order phase transition into a quantum Hall ferromagnetic
state,  whilst the coefficient $\tilde{b}_z$ acts as a 
symmetry-breaking field.  We include the transition temperatures $T_{QHF}$ obtained 
from the condition $a_{I} = 0$ in the numerical examples in Sec.~\ref{sec:phased} below.
 
\subsection{Phase Diagrams}
\label{sec:phased}
We study the phase diagram as a function of $\alpha$, $\nu_{T}$
and $T$ in two different directions in parameter space.  
First we vary $\nu_T$ from 7.5 to 8.0 when the energy levels are degenerate. 
Second, we vary $\alpha$ at fixed  $\nu_{T}=7.5$  
and 8.0.  It should be noted\cite{MoessnerChalker} that the free
energy expansion is strictly only valid when the order parameters are small
and is hence most applicable near the transition temperature into 
an ordered state. Thus, at a given $\alpha$ and $\nu_T$, we consider the 
ground state to be the state with the highest ordering temperature.

\subsubsection{Degenerate energy levels}
\label{sec:degen}

\begin{figure}
\includegraphics[width=3in]{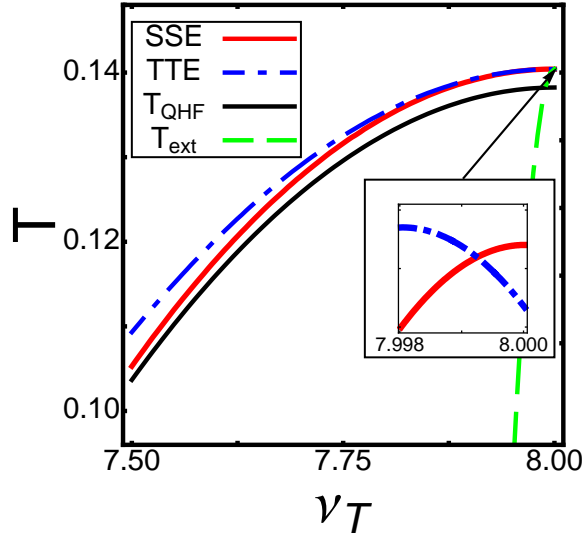}
\caption{\label{fig:PhaseNuTemp}Phase diagram in the
$\nu_{T}-T$ plane when energy levels are degenerate. Temperature
is in units $\frac{e^{2}}{4\pi\epsilon l_{0}k_{B}}$. $T_\mathrm{ext}$ is the extrapolation of the line separating the TTE and SSE phases to low temperature.~\cite{MoessnerChalker}
Inset: The SSE phase has a slightly higher ordering temperature than the TTE
phase in a very narrow range around $\nu_T=8.00$.}
\end{figure}

In Fig.~\ref{fig:PhaseNuTemp} we show the phase diagram as a function
of $\nu_T$ for the energy levels found in Sec.~\ref{sec:states} when the 
two states are degenerate.   As shown in the inset, when  $\nu_T  \simeq 8.0$, 
there is a transition from the uniform phase to the SSE phase. 
 As $\nu_T$ is decreased from 8.0,  the transition from uniform 
to CDW ordering prefers the TTE phase. Below the transition temperature, 
the transition between the SSE and TTE phases is first-order, 
and is a function of temperature and total filling fraction.  Note that 
the Quantum Hall ferromagnetism temperature $T_{QHF}$ is lower than the CDW 
ordering temperature for all $\nu_T$ considered.  Thermal and quantum 
fluctuations might change this ordering of transition temperatures.

For each of the mixed states we find numerically that the order parameters 
$\Delta_1$ and $\Delta_2$ do not have the same phase.  This can be understood by
considering the lowest order term in the free energy that contains $\Delta_1$
and $\Delta_2$ (see Appendix \ref{sec:Landau-free-energy-coefficients-q1=00003D=00003Dq2}):
 \begin{eqnarray} 
\delta \mathcal{F}_{\mathbf{Q}_1 =\mathbf{Q}_2}^{\left(2\right)} & = &
P_1 P_2 U_{12}\left(Q\right)\left[ 
1+U_{11}\left(Q\right)f'\left(E_{1}-\mu\right)
 +U_{22}\left(Q\right)f'\left(E_{2}-\mu\right)\right] 
|\Delta_{1}||\Delta_{2}|\cos\left(\delta\phi_{12}\right). 
\label{eq:delta1delta2}
\end{eqnarray}
Here, $\delta\phi_{12}$ is the relative phase between
the order parameters. The term in Eq.~(\ref{eq:delta1delta2}) changes sign at 
$T = T_{12}=\nu_{1}\left(1-\nu_{1}\right)U_{11}\left(Q_{12}\right)+\nu_{2}\left(1-\nu_{2}\right)U_{22}\left(Q_{12}\right)$.
Since $U_{12}\left(Q_{12}\right)$ is negative for the values of
$Q_{12}$ that minimize the free energy near the transition, the sign of this term
below $T_{12}$ depends on the sign of $\cos\left(\delta\phi_{12}\right)$.  In the 
SSE state, $\delta\phi_{12} = \pi$, while for the TTE state, 
$\phi_{2i}-\phi_{1i} = \pi$, $i=1$, 2, 3. Thus, the SSE and TTE CDW states 
can be thought of as coexisting CDW phases in $\Phi_1$ and $\Phi_2$ 
ordered at the same wavevector, but $\pi$ out of phase with one another.

It is possible to determine another characteristic temperature associated with equal
 wavevector ordering in addition to $T_{12}$. Truncating the free energy to 
$\mathcal{O}\left(\Delta_i^2\right)$ and minimizing with respect to both of the order 
parameters and their relative phases leads to the following quadratic equation whose solution yields a
characteristic temperature $T_m(Q)$:
\begin{eqnarray}
\label{eq:SecondOrderEqQTemp}
\left(1-\beta\right)T^2 - 2\left(T_{12}-\beta\frac{T_1+T_2}{2}\right)T + \left(T_{12}^2-\beta T_1 T_2\right) & = & 0,
\end{eqnarray}
where $T_1 = U_{11}\left(Q\right)\nu_1\left(1-\nu_1\right) + U_{12}\left(Q\right)^2/U_{11}\left(Q\right)\nu_2\left(1-\nu_2\right)$ 
(similar for $T_2$), $T_{12}$ is defined from Eq.~(\ref{eq:delta1delta2}), 
and $\beta = 4 U_{11}\left(Q\right) U_{22}\left(Q\right)/U_{12}\left(Q\right)^2$. 
Solving Eq.~(\ref{eq:SecondOrderEqQTemp}) for $T=T_{m}\left(Q\right)$ gives us an 
estimate of the transition temperature at which equal wavevector CDW ordering becomes 
relevant. Furthermore, maximizing $T_{m}\left(Q\right)$ with respect to $Q$ gives us 
the critical value of the wavevector at the transition temperature. 
When terms beyond Eq.~(\ref{eq:delta1delta2}) are included in the free energy,
it is not possible to obtain an analytical solution as we have above,
and a numerical solution is required. $T_{12}$ as defined above is
larger than the ordering temperatures we find numerically and the presence
of cubic terms in the free energy generically leads to a first order
transition to the TTE state.

For the particular example we consider here, at the ordering wavevector $Q$, 
$U_{11}(Q)/U_{22}(Q) \sim 2$, and we find that the ordering temperature for
TTE is marginally higher than the ordering temperature for T1 which is much
larger than the ordering temperature for T2.  The tendency towards ordering
is much stronger in state 1 than state 2, but the small energy gain from out
of phase equal wavevector ordering in the two states implies that that the 
TTE phase has the highest ordering temperature.

\subsubsection{Phase diagram at fixed $\nu_T$}
\label{sec:nondegen}

\begin{figure}[htb]
\includegraphics[width=3in]{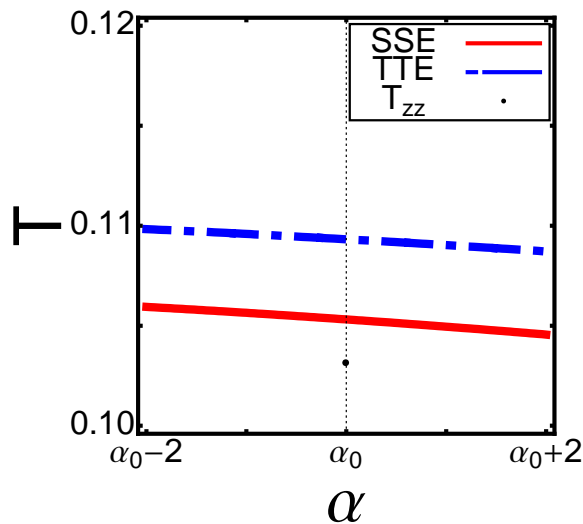}\caption{\label{fig:PhaseAlpha750Colour}Phase diagram 
at $\nu_{T}=7.5$ in the $\alpha-T$ plane. Temperature is in units
of $\frac{e^{2}}{4\pi\epsilon l_{0}k_{B}}$ and $\alpha$ is dimensionless.
The degeneracy point is indicated by a dotted black line on the $\alpha$ axis.}
\end{figure}

\begin{figure}
\includegraphics[width=3in]{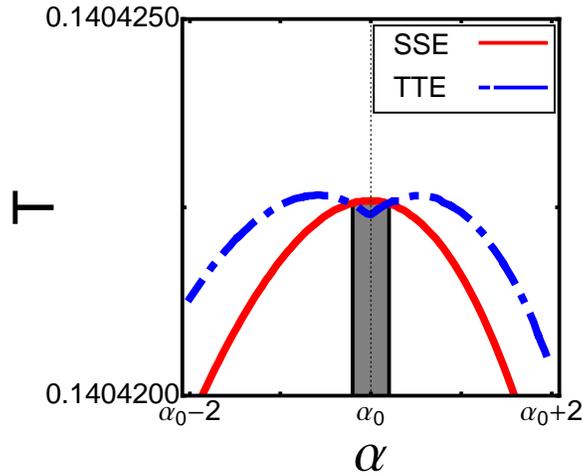}\caption{\label{fig:PhaseAlpha800Colour}Phase diagram 
at $\nu_{T}=8.0$ in the $\alpha-T$ plane. Temperature is in units
of $\frac{e^{2}}{4\pi\epsilon l_{0}k_{B}}$ and $\alpha$ is dimensionless.
The degeneracy point is indicated by a dotted black line on the $\alpha$ axis.  
The shaded region indicates where SSE ordering is preferred over TTE ordering. Temperature $T_{zz}$ is below the temperature range.}
\end{figure}

An orthogonal trajectory in phase space to that considered in Sec.~\ref{sec:degen}
is to fix $\nu_T$ and then vary $\alpha$ and $T$. As $\alpha$ is varied, the filling fractions
of the two states vary,  which allows for phase transitions between
CDW states at fixed temperature.  Away from 
degeneracy, temperature also tunes the relative filling of the two states,
with the filling of the higher energy state going to zero at low temperature.
We plot phase diagrams for $\nu_T = 7.50$ and 8.00.  These are
shown in Figs.~\ref{fig:PhaseAlpha750Colour} 
and \ref{fig:PhaseAlpha800Colour} respectively.

 At $\nu_{T}=7.50,$ we see
that the TTE phase dominates around the degeneracy point. At $\nu_{T}=8.00$, the SSE state is favoured at values of $\alpha$
close to the degeneracy point, and the TTE phase is favoured away from degeneracy and near the transition temperature.
Transitions between triangular and striped
phases are first-order in nature. Away from degeneracy and at low enough temperatures, we expect only the state 
with lowest energy to have non-zero filling, leading to an integer quantum Hall state. 
The quantum Hall ferromagnetic ordering temperature is below the CDW ordering temperatures
in both cases.  Similarly to the degenerate case, ordering in state 1 appears to be
the main driver of the transition with ordering in state 2 contributing a small 
lowering of the free energy.  The balance of the contribution to ordering of the two
states is non-generic and will depend on the details of the two non-interacting
states that are participating in the ordering.

\subsubsection{Generic features of the phase diagram}
\label{sec:Genfeatures}
In our numerical example, 
we have focused on a particular pair of levels crossing at a particular
filling fraction, so it is natural to ask to what extent the features we observe in 
the phase diagram are generic to when there is Landau level mixing as
opposed to specific to the two individual levels that we selected. 
In general, the choice of two particular levels will affect the coefficients in
the Landau theory through
the mixing coefficients $C_{\Phi\alpha}$ (as introduced in 
Eq.~(\ref{eq:Eigenstates})). 
These in turn determine the allowed phases, the 
ordering temperatures and associated wavevectors, 
and the relative magnitudes of the relevant order parameters in mixed states. We 
considered several other crossings in less detail than the one presented here, and
these investigations, combined with general expectations, allow us to outline the
features we believe to be more or less generic for CDW ordering when there are two
levels close to degeneracy.
The $\nu_T$-$T$ phase diagram we show in Fig.~\ref{fig:PhaseNuTemp} is 
qualitatively very similar to the temperature-filling phase diagrams found 
in the case of a {\emph single} LL by Fogler {\it et al.} 
(FKS)\cite{FKS} and Moessner and Chalker (MC)\cite{MoessnerChalker}
if we divide $\nu_0 = \nu_1 + \nu_2$ by 2.  In both of these works it
was found that for $\nu$ close to $\frac{1}{2}$ in the valence LL,
there is a striped CDW, with a transition to a triangular CDW at zero temperature for
$\nu_c \gtrsim 0.35$ (FKS) or $\nu_c \gtrsim 0.39$ (MC). Taking a similar approach, 
we extrapolate the line separating the SSE and TTE phases 
in Fig.~\ref{fig:PhaseNuTemp} to zero temperature 
and find that the SSE phase is stable 
when the effective filling fraction of each state is approximately
$\nu_c \gtrsim 0.45$.  As MC point out, the perturbative expansion of the
free energy is not valid at very low temperatures, so our estimate of 
$\nu_c$ is only approximate. Additionally, we can reasonably expect that the
precise estimate of $\nu_c$ may depend on which two levels are being studied.
In this sense, the results here for degenerate energy levels 
are consistent with known results on single-LL systems.  However, there is 
an important difference, in that there are effectively two half-filled
states and there is co-ordinated ordering in both of them, 
as opposed to the single half-filled Landau level studied by FKS and
MC.  

Away from degeneracy the TTE  phase tends to be dominant.
As $\nu_0$ increases towards unity, the SSE phase has a higher transition 
temperature just near degeneracy.  In all of the phases where
there is ordering in both states 1 and 2, we observed that the wavevectors $\bvec{Q}_1
 = \bvec{Q}_2$, which can be understood as arising from the extra free energy that
can be gained when the wavevectors are equal in the two states and the order 
parameters are in phase, as illustrated in Eq.~(\ref{eq:delta1delta2}). 

At degeneracy, we compared the transition temperatures of the CDW states 
to the QHF states, and found that the system preferred CDW 
ordering over the QHF ordering.  We are uncertain as to the extent to which 
this behaviour is generic, however, our results appear to indicate that the CDW state
can be competitive with quantum Hall ferromagnetism.

\section{Discussion and Conclusions}
\label{sec:Conclusion}

We have considered charge density wave ordering in a 2D hole system in a 
perpendicular magnetic field when Landau level mixing is important.
The motivation for this work is recent transport experiments in
 quantum Hall systems where Landau level mixing was induced 
by spin-orbit coupling.\cite{Manfra,Grayson} 
Our work is related to the situations investigated by
Manfra \emph{et al.}\cite{Manfra} 
and Fischer {\it et al.}\cite{Grayson} There has been some previous theoretical 
work on both hole striped states\cite{Kim} and hole FQHE states,\cite{Yanghole} 
but these works do not focus on the situation when two levels are close 
to degenerate and the possible orderings in such a case, as we do here.

Recently, Manfra \emph{et al.}\cite{Manfra} studied the transport
properties of a two-dimensional hole system in a perpendicular magnetic
field. They found anisotropic resistivity at some, but not all, half-integer
filling fractions; in particular, they observed isotropic charge transport
at $\nu$=9/2, flanked by anisotropic transport at $\nu$=7/2 and
$\nu$=11/2. They argue that LL mixing caused by spin-orbit coupling
in the valence band of GaAs may be responsible for this pattern of
isotropic and anisotropic transport. From a self-consistent calculation
of the Landau levels in the Hartree approximation, they suggested that
the contribution from the $N=0$ and $N=1$ LLs to the wavefunction of the $\nu=9/2$
state in their particular device likely suppresses the tendency towards
charge density wave order.  

In Fig.~\ref{fig:PhaseAlpha750Colour}
we study the situation in which $\nu_0 = 0.5$ and find that well away from
degeneracy so long as there is some occupation of the upper energy level  
$\nu_1 \gg \nu_2 \neq 0$ the effect of LL mixing 
can be to stabilize a mixed triangular CDW state over 
the striped CDW state we would expect for an unmixed Landau level.  This is
another mechanism that could lead to the absence of anisotropic transport 
at $\nu_0 = 0.5$.

Fischer {\it et al.}\cite{Grayson} observed anomalous transport in the lowest 
Landau level at what they interpreted to be a field-induced
level anti-crossing.  At a temperature of 320 mK there was a peak
in the longitudinal resistivity that disappeared at  50 mK.  
 As we do not expect our theory to be
applicable to the lowest Landau level, we do not speculate on
the origin of this anomalous transport.  However, we do note that 
their experiment exhibits a level anti-crossing that occurs 
because they study a two-dimensional hole rather than a two 
dimensional electron system.  
Rashba and Dresselhaus coupling play an important role in 
determining the energy levels in their device, and
 if they tuned the Rashba coupling,  they would
be able to explore the effects of tuning mixing at $\nu_0 = 0.5$ and 
$\nu_0 = 1$ directly.

Our calculations generalize existing work on 2DEGs in high-LLs to the
situation when LL mixing is important, and we show that for a contact 
potential  the Hartree-Fock approximation is exact in the high LL limit
even in the presence of LL mixing. We specialize to the situation in which 
 two levels are near the Fermi energy and perform a diagrammatic
analysis of the two-particle vertex.  This leads us to predict 
three relevant temperature scales where the uniform system becomes unstable 
to charge-density wave formation. Two of these correspond to CDW formation 
in either of the two levels, and the third corresponds to an 
instability to a mixed CDW phase. 
Having established a tendency towards CDW ordering, we derive a Landau
theory for CDW ordering in the two levels using the Hartree-Fock 
approximation.  We then focus on the specific example of LL mixing
induced by Rashba spin-orbit coupling (which could provide a tuning parameter
for experimental investigations of the effects of LL mixing on anisotropic transport
in the quantum Hall regime).

The effect of Landau level mixing goes beyond
changing the character of the single-particle states in the system to
also changing the spectrum and the spacing in energy of the single-particle
states.  We find that it is the second of these that appears to have the
most effect on anisotropic quantum Hall states, once the single-particle
states have little $N=0$ or 1 character.  In particular, at the level
of Hartree-Fock, if there are two states relatively near the Fermi energy, 
only when the filling of one or both states is very close
to $\nu = 0.5$ is there striped CDW ordering.  At other fillings there is 
triangular CDW ordering.  A competing phase when there are two close to 
degenerate states is that of quantum Hall ferromagnetism, and while in the 
example we consider here we find the ordering temperature to be less than 
that for a CDW, thermal fluctuations may treat the two states differently.

We hope that our work stimulates further 
theoretical and experimental study of anisotropic Quantum Hall states 
in which externally tuned LL mixing
is used as a parameter to investigate the phase diagram.

\section{Acknowledgements}
The authors thank Igor Herbut and Kun Yang for helpful discussions.
This work was supported by NSERC.

\appendix

\section{Full expressions for coefficients in Landau theory}

\label{sec:Landau-free-energy-coefficients}

In this Appendix, we give exact expressions for the coefficients in
the Landau free energy theory for each case discussed in Sec.~\ref{sec:Landau-free-energy-theory}
when $\bvec{Q}_{1}\neq\bvec{Q}_{2}$. In solving for the $c_{i}$
and $\gamma$ terms, we determine the chemical potential by setting the
number of fermions in the non-interacting system equal to the number
of fermions in the interacting system. Then, \begin{eqnarray}
\mu-\mu_{0} & = & -\frac{1}{T}\frac{\sum_{kk'}\sum_{\left\{ \mathbf{Q}_{k}\right\} }P_k^2\left|U_{kk'}\right|^{2}\nu_{k'}\left(1-\nu_{k'}\right)\left(1-2\nu_{k'}\right)\Delta_{k}^{2}}{\sum_{k}\nu_{k}\left(1-\nu_{k}\right)}.\end{eqnarray}

\subsection{Mixed unidirectional-unidirectional CDW system}

The general form of the free energy for striped phases is 
\begin{eqnarray}
\mathcal{F} & = &  P_1^2 a_{1}\left(\mathbf{Q}_{1},T\right)\left|\Delta_{1}\right|^{2}
                           + P_2^2 a_{2}\left(\mathbf{Q}_{2},T\right)\left|\Delta_{2}\right|^{2} 
                           + P_1^4 c_{1}\left(\mathbf{Q}_{1},T\right)\left|\Delta_{1}\right|^{4} 
                           + P_2^4 c_{2}\left(\mathbf{Q}_{2},T\right)\left|\Delta_{2}\right|^{4}\nonumber\\
                   &    & + P_1^2 P_2^2 \gamma\left(\mathbf{Q}_{1},\mathbf{Q}_{2},\theta,T\right)\left|\Delta_{1}\right|^{2}\left|\Delta_{2}\right|^{2},\label{eq:StripeStripeFreeEnergy}
\end{eqnarray}
Note that terms involving $\Delta_i^3$ are not allowed by symmetry.
 The parameters for the free energy theory in the case of mixed unidirectional-unidirectional
CDW ordering are
\begin{eqnarray}
a_{iS}\left(\mathbf{Q}_{i},T\right) & = & U_{ii}\left(Q_{i}\right)+U_{ii}\left(Q_{i}\right)^{2}f'\left(E_{i}-\mu\right)+\sum_{j\neq i}U_{ij}\left(Q_{i}\right)^{2}f'\left(E_{j}-\mu\right),
\end{eqnarray}

\begin{eqnarray}
c_{iS}\left(\mathbf{Q}_{i},T\right) & = & \frac{1}{4}\Biggl[U_{ii}\left(Q_{i}\right)^{4}f'''\left(E_{i}-\mu\right)+\sum_{j\neq i}U_{ij}\left(Q_{i}\right)^{4}f'''\left(E_{j}-\mu\right)\Biggr]\nonumber\\
 &  & -\frac{2}{\sum_{k}f'\left(E_{k}-\mu\right)}\biggl\{U_{ii}\left(Q_{i}\right)^{4}f''\left(E_{i}-\mu\right)^{2}
%\nonumber\\  &  & \qquad
+\sum_{j\neq i}\bigl[ U_{ij}\left(Q_{i}\right)^{4}f''\left(E_{j}-\mu\right)^{2} \nonumber \\
 &  & \qquad + \, 2U_{ii}\left(Q_{i}\right)^{2}U_{ij}\left(Q_{j}\right)f''\left(E_{i}-\mu\right)f''\left(E_{j}-\mu\right)\bigr]\biggr\},
 \end{eqnarray}
 and 
 \begin{eqnarray}
\gamma_{SS}\left(\mathbf{Q}_{1},\mathbf{Q}_{2},T\right) & = & \frac{1}{3}U_{11}\left(Q_{1}\right)U_{12}\left(Q_{2}\right)^{2}f'''\left(E_{1}-\mu\right)\biggl[2+\cos\left(\tilde{Q}_{1}Q_{2}\sin\theta\right)\biggr]\nonumber\\
 &  & +\frac{1}{3}U_{12}\left(Q_{1}\right)^2U_{22}\left(Q_{2}\right)^{2}f'''\left(E_{2}-\mu\right)\biggl[2+\cos\left(Q_{1}Q_{2}\sin\theta\right)\biggr]\nonumber\\
 &  & -\frac{1}{\sum_{i}f'\left(E_{i}-\mu\right)}\biggl[f''\left(E_{1}-\mu\right)f''\left(E_{2}-\mu\right)U_{11}\left(Q_{1}\right)^{2}U_{22}\left(Q_{2}\right)^{2}\nonumber\\
 &  & \qquad\qquad + f''\left(E_{1}-\mu\right)^{2}U_{11}\left(Q_{1}\right)^{2}U_{12}\left(Q_{2}\right)^{2}%
%\nonumber\\ &  & \qquad\qquad 
+ f''\left(E_{2}-\mu\right)^{2}U_{12}\left(Q_{1}\right)^{2}U_{22}\left(Q_{2}\right)^{2}\nonumber\\
 &  & \qquad\qquad + f''\left(E_{1}-\mu\right)f''\left(E_{2}-\mu\right)U_{12}\left(Q_{1}\right)^{2}U_{21}\left(Q_{2}\right)^{2}\biggr].
\end{eqnarray}

\subsection{Mixed triangular-triangular CDW}

The free energy when there is mixed triangular-triangular ordering is 
\begin{eqnarray}
\mathcal{F} & = &  P_1^2 a_{1}\left(\mathbf{Q}_{1},T\right)\left|\Delta_{1}\right|^{2}
                           + P_2^2 a_{2}\left(\mathbf{Q}_{2},T\right)\left|\Delta_{2}\right|^{2} 
                           + P_1^3 b_{1}\left(\mathbf{Q}_{1},T\right)\left|\Delta_{1}\right|^{3} 
                           + P_2^3 b_{2}\left(\mathbf{Q}_{1},T\right)\left|\Delta_{2}\right|^{3}\nonumber \\
                 &    & + \hspace{0.1cm}P_1^4 c_{1}\left(\mathbf{Q}_{1},T\right)\left|\Delta_{1}\right|^{4} 
                           + P_2^4 c_{2}\left(\mathbf{Q}_{2},T\right)\left|\Delta_{2}\right|^{4} 
                           + P_1^2 P_2^2 \gamma\left(\mathbf{Q}_{1},\mathbf{Q}_{2},\theta,T\right)\left|\Delta_{1}\right|^{2}\left|\Delta_{2}\right|^{2},\label{eq:TriangTriangFreeEnergy-1}
\end{eqnarray}
 A triangular CDW is described by a set of three wavevectors $\left\{ \mathbf{Q}_{k}^{j}\right\} _{j=1}^{3}$
lying on a circle of radius $Q_{k}$ oriented $\frac{2\pi}{3}$ radians
apart. Allowing for non-collinear CDW ordering between two triangular
CDW states, we define $\theta$ as the angle between wavevectors $\mathbf{Q}_{1}^{1}$
and $\mathbf{Q}_{2}^{1}$. Then, the angle between any two wavevectors
in $\Phi_{1}$ and $\Phi_{2}$ is $\theta\left(\mathbf{Q}_{1}^{j},\mathbf{Q}_{2}^{j'}\right)=\theta+\frac{2\pi}{3}\left(j'-j\right).$
There are three possible angles between wavevectors: $\left\{ \theta_{i}\right\} _{i=1}^{3}=\left\{ \theta,\theta+\frac{2\pi}{3},\theta-\frac{2\pi}{3}\right\} $.

The parameters in this free energy model are 
\begin{eqnarray}
a_{iT}\left(\mathbf{Q}_{i},T\right) 
& = & 3\Biggl[U_{ii}\left(Q_{i}\right)+U_{ii}\left(Q_{i}\right)^{2}f'\left(E_{i}-\mu\right)+\sum_{j\neq i}U_{ij}\left(Q_i\right)^{2}f'\left(E_{j}-\mu\right)\Biggr],
%\nonumber \\ &    &
\end{eqnarray}
 \begin{eqnarray}
b_{i}\left(\mathbf{Q}_{i},T\right) 
& = & -2\cos\left(\frac{\sqrt{3}Q_{i}^{2}}{4}\right)\Biggl[U_{ii}\left(Q_{i}\right)^{3}f^{\prime\prime}\left(E_{i}-\mu\right)+\sum_{j\neq i}U_{ij}\left(Q_{i}\right)^{3}f^{\prime\prime}\left(E_{j}-\mu\right)\Biggr],
%\nonumber \\ &    &
\end{eqnarray}
\begin{eqnarray}
c_{iT}\left(\mathbf{Q}_{i},T\right) & = & \frac{15}{4}U_{ii}\left(Q_{i}\right)^{4}f'''\left(E_{i}-\mu\right)-U_{ii}\left(Q_{i}\right)^{4}f'''\left(E_{i}-\mu\right)\Biggl[1-\cos\left(\frac{\sqrt{3}}{2}Q_{i}^{2}\right)\Biggr]\nonumber\\
 &  & +\sum_{j\neq i}\Biggl\{\frac{15}{4}U_{ij}\left(Q_{i}\right)^{4}f'''\left(E_{j}-\mu\right)
%\nonumber\\  &   &\qquad 
- U_{ij}\left(Q_{i}\right)^{4}f'''\left(E_{j}-\mu\right)\Biggl[1-\cos\left(\frac{\sqrt{3}}{2}Q_{i}^{2}\right)\Biggr]\Biggr\}\nonumber\\
 &  & - \frac{9}{2\sum_{i}f'\left(E_{i}-\mu\right)}\biggl\{ U_{ii}\left(Q_{i}\right)^{4}f''\left(E_{i}-\mu\right)^{2}\nonumber\\
 &  & \qquad+\sum_{j\neq i}\biggl[f''\left(E_{j}-\mu\right)^{2}U_{ij}\left(Q_{i}\right) 
%\nonumber \\  &  & \qquad 
+ 2f''\left(E_{i}-\mu\right)f''\left(E_{j}-\mu\right)U_{ii}\left(Q_{i}\right)^{2}U_{ij}\left(Q_{i}\right)\biggr]\biggr\},\end{eqnarray}
 and 
 \begin{eqnarray}
\gamma_{TT}\left(\mathbf{Q}_{1},\mathbf{Q}_{2},T\right) 
& = & \left[U_{11}\left(Q_{1}\right)^{2}U_{12}\left(Q_{2}\right)^{2}f'''\left(E_{1}-\mu\right) 
%\right. \nonumber \\ &    &\qquad \left.
+ U_{11}\left(Q_{1}\right)^{2}U_{12}\left(Q_{2}\right)^{2}f'''\left(E_{1}-\mu\right)\right]\nonumber\\
&    & \times\biggl\{6+\cos\left(Q_{1}Q_{2}\sin\theta\right)%\nonumber\\
%&    & \qquad 
+\cos\left[Q_{1}Q_{2}\sin\left(\theta+\frac{2\pi}{3}\right)\right]+\cos\left[Q_{1}Q_{2}\sin\left(\theta-\frac{2\pi}{3}\right)\right]\biggr\}\nonumber\\
&    & -\frac{9}{\sum_{i}f'\left(E_{i}-\mu\right)}\biggl[f''\left(E_{1}-\mu\right)f''\left(E_{2}-\mu\right)U_{11}\left(Q_{1}\right)^{2}U_{22}\left(Q_{2}\right)^{2}\nonumber\\
&    & \qquad+f''\left(E_{1}-\mu\right)^{2}U_{11}\left(Q_{1}\right)^{2}U_{12}\left(Q_{2}\right)^{2} %\nonumber\\
%&    & \qquad
+f''\left(E_{2}-\mu\right)^{2}U_{12}\left(Q_{1}\right)^{2}U_{22}\left(Q_{2}\right)^{2}\nonumber\\
&    & \qquad+f''\left(E_{1}-\mu\right)f''\left(E_{2}-\mu\right)U_{12}\left(Q_{1}\right)^{2}U_{22}\left(Q_{2}\right)^{2}\biggr].
\end{eqnarray}
 
\subsection{Mixed unidirectional-triangular CDW}

Without loss of generality, assume that there is unidirectional CDW
ordering in $\Phi_{1}$ and triangular CDW ordering in $\Phi_{2}$.
Then the free energy  is 

\begin{eqnarray}
\mathcal{F} & = & P_1^2 a_{1}\left(\mathbf{Q}_{1},T\right)\left|\Delta_{1}\right|^{2} 
                          + P_2^2 a_{2}\left(\mathbf{Q}_{2},T\right)\left|\Delta_{2}\right|^{2} 
                          + P_2^3 b_{2}\left(\mathbf{Q}_{1},T\right)\left|\Delta_{2}\right|^{3} \nonumber\\
                   &    & + P_1^4 c_{1}\left(\mathbf{Q}_{1},T\right)\left|\Delta_{1}\right|^{4} 
                          + P_2^4 c_{2}\left(\mathbf{Q}_{2},T\right)\left|\Delta_{2}\right|^{4}
                          + P_1^2 P_2^2 \hspace{0.1cm} \gamma\left(\mathbf{Q}_{1},\mathbf{Q}_{2},\theta,T\right)\left|\Delta_{1}\right|^{2}\left|\Delta_{2}\right|^{2},\label{eq:StripeTriangFreeEnergy}
\end{eqnarray}
 All coefficients except $\gamma$ remain unchanged for each type
of ordering. The coupling parameter is 
\begin{eqnarray}
\gamma_{ST}\left(\mathbf{Q}_{1},\mathbf{Q}_{2},T\right) 
& = & \frac{1}{3}\left[U_{11}\left(Q_{1}\right)^2U_{12}\left(Q_{2}\right)^2f'''\left(E_{1}-\mu\right) \right.+\left.U_{11}\left(Q_{1}\right)^2U_{12}\left(Q_{2}\right)^2f'''\left(E_{1}-\mu\right)\right]\nonumber \\
 &  & \qquad\times\biggl\{6+\cos\left(Q_{1}Q_{2}\sin\theta\right) +\cos\left[Q_{1}Q_{2}\sin\left(\theta+\frac{2\pi}{3}\right)\right]+\cos\left[Q_{1}Q_{2}\sin\left(\theta-\frac{2\pi}{3}\right)\right]\biggr\}\nonumber\\
 &  & -\frac{3}{\sum_{i}f'\left(E_{i}-\mu\right)}\biggl[f''\left(E_{1}-\mu\right)f''\left(E_{2}-\mu\right)U_{11}\left(Q_{1}\right)^{2}U_{22}\left(Q_{2}\right)^{2}\nonumber\\
 &  & \qquad+f''\left(E_{1}-\mu\right)^{2}U_{11}\left(Q_{1}\right)^{2}U_{12}\left(Q_{2}\right)^{2} +f''\left(E_{2}-\mu\right)^{2}U_{12}\left(Q_{1}\right)^{2}U_{22}\left(Q_{2}\right)^{2}\nonumber\\
 &  & \qquad+f''\left(E_{1}-\mu\right)f''\left(E_{2}-\mu\right)U_{12}\left(Q_{1}\right)^{2}U_{22}\left(Q_{2}\right)^{2}\biggr].
 \end{eqnarray}

\section{Full expressions for coefficients in Landau theory when $\bvec{Q}_{1}=\bvec{Q}_{2}$}

\label{sec:Landau-free-energy-coefficients-q1=00003D=00003Dq2} In
the previous appendix we gave exact expressions for the coefficients in the
Landau free energy theory for each case discussed in Sec.~\ref{sec:Landau-free-energy-theory}.  When the ordering wavevectors $\bvec{Q}_{1}=\bvec{Q}_{2}=\bvec{Q}$
then there are additional terms in the free energy, as described in 
Eq.~(\ref{eq:deltafree}) as $\delta{\mathcal{F}}_{\bvec{Q}_{1}=\bvec{Q}_{2}}$.  
We give expressions for the coefficients in 
Eq.~(\ref{eq:deltafree}) in this Appendix. 

\subsection{Mixed unidirectional-unidirectional CDW system}

The extra terms that appear in the free energy theory for SSE ordering
are of the form 
\begin{eqnarray}
\delta f_{SSE} & = & P_1 P_2\alpha_{SSE}\left(Q,\delta\phi,T\right)\left|\Delta_{1}\right|\left|\Delta_{2}\right| 
                                + P_1^3 P_2\sigma_{SSE}\left(Q,\delta\phi,T\right)\left|\Delta_{1}\right|^{3}\left|\Delta_{2}\right| \nonumber \\
                       &    & + P_1^2 P_2^2\gamma_{SSE}\left(Q,\delta\phi,T\right)\left|\Delta_{1}\right|^{2}\left|\Delta_{2}\right|^{2}
                                + P_1 P_2^3\rho_{SSE}\left(Q,\delta\phi,T\right)\left|\Delta_{1}\right|\left|\Delta_{2}\right|^{3}.
\end{eqnarray}
 Here, $Q$ is the magnitude of the ordering wavevector, $\delta\phi$
the relative phase between order parameters $\Delta_{1}\left(\mathbf{Q}\right)$
and $\Delta_{2}\left(\mathbf{Q}\right)$, and $T$ the temperature.
The values of the coefficients are 
\begin{eqnarray}
\alpha_{SSE}\left(Q,\delta\phi,T\right) & = & 2U_{12}\left(Q\right) 
%\nonumber\\
%&  & \times
\biggl[1+U_{11}\left(Q\right)f'\left(E_{1}-\mu\right)+U_{22}\left(Q\right)f'\left(E_{2}-\mu\right)\biggr]\cos\left(\delta\phi\right),
\end{eqnarray}
 \begin{eqnarray}
\sigma_{SSE}\left(Q,\delta\phi,T\right) & = & \biggl\{ U_{11}\left(Q\right)^{3}U_{12}\left(Q\right)f'''\left(E_{1}-\mu\right)+U_{12}\left(Q\right)^{3}U_{11}\left(Q\right)f'''\left(E_{2}-\mu\right)\nonumber\\
 &  & \qquad-\frac{1}{\sum_{i=1}^{2}f'\left(E_{i}-\mu\right)}\left[U_{11}\left(Q\right)^{2}f''\left(E_{1}-\mu\right)+U_{12}\left(Q\right)^{2}f''\left(E_{2}-\mu\right)\right]\nonumber\\
 &  & \qquad\times\left[U_{11}\left(Q\right)U_{12}\left(Q\right)f''\left(E_{1}-\mu\right)+U_{22}\left(Q\right)U_{12}\left(Q\right)f''\left(E_{2}-\mu\right)\right]\biggr\} %\nonumber\\
% &  &\times
\cos\left(\delta\phi\right),
\end{eqnarray}
\begin{eqnarray}
\gamma_{SSE}\left(Q,\delta\phi,T\right) 
& = & \biggl\{\frac{1}{2}\biggl[U_{11}\left(Q\right)^{2}U_{12}\left(Q\right)^{2}f'''\left(E_{1}-\mu\right)+U_{12}\left(Q\right)^{2}U_{22}\left(Q\right)^{2}f'''\left(E_{2}-\mu\right)\biggr]\nonumber\\
&    & \qquad-\frac{1}{\sum_{i=1}^{2}f'\left(E_{i}-\mu\right)}\nonumber\\
&    & \qquad\times \left[U_{11}\left(Q\right)U_{12}\left(Q\right)f''\left(E_{1}-\mu\right)+U_{22}\left(Q\right)U_{12}\left(Q\right)f''\left(E_{2}-\mu\right)\right]^{2}\biggr\}%\nonumber\\
% &   &\times
\left[1+2\cos^{2}\left(\delta\phi\right)\right]\nonumber\\
 &  & -\frac{1}{\sum_{i=1}^{2}f'\left(E_{i}-\mu\right)}
\left[U_{11}\left(Q\right)U_{22}\left(Q\right)-U_{12}\left(Q\right)^{2}\right]^{2}f''\left(E_{1}-\mu\right)f''\left(E_{2}-\mu\right),
 \end{eqnarray}
 and
\begin{eqnarray}
\rho_{SSE}\left(Q,\delta\phi,T\right) & = & \biggl\{ U_{11}\left(Q\right)U_{12}\left(Q\right)^{3}f'''\left(E_{1}-\mu\right)+U_{12}\left(Q\right)U_{22}\left(Q\right)^{3}f'''\left(E_{2}-\mu\right)\nonumber\\
 &  & \qquad-\frac{1}{\sum_{i=1}^{2}f'\left(E_{i}-\mu\right)}\left[U_{22}\left(Q\right)^{2}f''\left(E_{2}-\mu\right)+U_{12}\left(Q\right)^{2}f''\left(E_{1}-\mu\right)\right]\nonumber\\
 &  & \qquad\times\left[U_{11}\left(Q\right)U_{12}\left(Q\right)f''\left(E_{1}-\mu\right)+U_{22}\left(Q\right)U_{12}\left(Q\right)f''\left(E_{2}-\mu\right)\right]^{2}\biggr\}%\nonumber\\
% &  & \times 
\cos\left(\delta\phi\right).
 \end{eqnarray}

\subsection{Mixed triangular-triangular CDW system}

The order parameter for the triangular CDW state depends on a set
of three wavevectors, $\left\{ \mathbf{Q}_{i}^{\alpha}\right\} _{\alpha=1}^{3}$.
We associate ordering at each wavevector with a particular phase $\phi_{i}^{\alpha}$,
such that $\Delta_{i}\left(\mathbf{Q}_{i}^{\alpha}\right)=\Delta_{i}e^{i\phi_{\iota}^{\alpha}}$.
Thus, when $\mathbf{Q}_{1}=\mathbf{Q}_{2}$, we end up with six new
parameters to consider when minimizing the free energy. We may reduce
the number of parameters from six to four in the following manner (where $i = 1,\,2,\,3$):
\begin{eqnarray}
\delta\phi_{i} & = & \phi_{1}^{i}-\phi_{2}^{i} , \\
\tilde{\phi}_{1} & = & \sum_{i=1}^{3}\phi_{1}^{i},\\
\tilde{\phi}_{2} & = & \tilde{\phi}_{1}-\sum_{i=1}^{3}\delta\phi_{i}
\end{eqnarray}
 The new terms in the free energy theory for TTE ordering are

\begin{eqnarray} 
\delta f_{TTE} & = &
P_1 P_2 \alpha_{TTE}\left(Q,\delta\phi_{1},\delta\phi_{2},\delta\phi_{3},T\right)\left|\Delta_{1}\right|\left|\Delta_{2}\right| 
+ P_1^2 P_2 \eta_{TTE}\left(Q,\delta\phi_{1},\delta\phi_{2},\delta\phi_{3},\tilde{\phi}_{1},T\right)\left|\Delta_{1}\right|^{2}\left|\Delta_{2}\right|\nonumber\\
& & + P_1 P_2^2 \varphi_{TTE}\left(Q,\delta\phi_{1},\delta\phi_{2},\delta\phi_{3},\tilde{\phi}_{2},T\right)\left|\Delta_{1}\right|\left|\Delta_{2}\right|^{2} 
+ P_1^3 P_2 \sigma_{TTE}\left(Q,\delta\phi_{1},\delta\phi_{2},\delta\phi_{3},T\right)\left|\Delta_{1}\right|^{3}\left|\Delta_{2}\right|\nonumber\\
&  & +P_1^2 P_2^2  \gamma_{TTE}\left(Q,\delta\phi_{1},\delta\phi_{2},\delta\phi_{3},T\right)\left|\Delta_{1}\right|^{2}\left|\Delta_{2}\right|^{2}
+P_1 P_2^3 \rho_{TTE}\left(Q,\delta\phi_{1},\delta\phi_{2},\delta\phi_{3},T\right)\left|\Delta_{1}\right|\left|\Delta_{2}\right|^{3}.\end{eqnarray}
 Here, $Q$ is the magnitude of the ordering wavevector, $\delta\phi$
the relative phase between order parameters $\Delta_{1}\left(\mathbf{Q}\right)$
and $\Delta_{2}\left(\mathbf{Q}\right)$, and $T$ the temperature.
The values of the coefficients are 
\begin{eqnarray}
\alpha_{TTE}\left(Q,\delta\phi_{1},\delta\phi_{2},\delta\phi_{3},T\right) 
& = & U_{12}\left(Q\right)\biggl[1+U_{11}\left(Q\right)f'\left(E_{1}-\mu\right)%\nonumber\\
%&    &\qquad
+U_{22}\left(Q\right)f'\left(E_{2}-\mu\right)\biggr]\sum_{i=1}^{3}\cos\left(\delta\phi_{i}\right),
\end{eqnarray}

\begin{eqnarray}
\eta_{TTE}\left(Q,\delta\phi_{1},\delta\phi_{2},\delta\phi_{3},\tilde{\phi}_{1},T\right) 
& = & -6\biggl[ U_{11}\left(Q\right)^{2}U_{12}\left(Q\right)f''\left(E_{1}-\mu\right) %\nonumber\\
%&    &   \qquad
+ U_{11}\left(Q\right)U_{12}\left(Q\right)^{2}f''\left(E_{2}-\mu\right)\biggr]\nonumber\\
&    &   \times\cos\left(\frac{\sqrt{3}}{4}Q^{2}\right)\sum_{i=1}^{3}\cos\left(\tilde{\phi}_{1}-\delta\phi_{i}\right),
 \end{eqnarray}
 
 \begin{eqnarray}
\varphi_{TTE}\left(Q,\delta\phi_{1},\delta\phi_{2},\delta\phi_{3},\tilde{\phi}_{1},T\right) & = & -6\biggl[ U_{12}\left(Q\right)^{2}U_{11}\left(Q\right)f''\left(E_{1}-\mu\right)%\nonumber\\
% &  & \qquad
+U_{12}\left(Q\right)U_{22}\left(Q\right)^{2}f''\left(E_{2}-\mu\right)\biggr]\nonumber\\
 &   & \times\cos\left(\frac{\sqrt{3}}{4}Q^{2}\right)\sum_{i=1}^{3}\cos\left(\tilde{\phi}_{2}+\delta\phi_{i}\right),
 \end{eqnarray}
 
\begin{eqnarray}
\sigma_{TTE}\left(Q,\delta\phi,T\right) 
& = & \biggl\{\frac{1}{3}\left[11+4\cos\left(\frac{\sqrt{3}}{2}Q^{2}\right)\right] 
%\nonumber\\ &   & \qquad\times
\left[U_{11}\left(Q\right)^{3}U_{12}\left(Q\right)f'''\left(E_{1}-\mu\right) + U_{22}\left(Q\right)U_{12}\left(Q\right)^{3}f'''\left(E_{2}-\mu\right)\right]\nonumber\\
&  & \qquad-\frac{6}{\sum_{i=1}^{2}f'\left(E_{i}-\mu\right)}
% \nonumber\\ &   & \qquad \times
\left[U_{11}\left(Q\right)U_{12}\left(Q\right)f''\left(E_{1}-\mu\right)+U_{22}\left(Q\right)U_{12}\left(Q\right)f''\left(E_{2}-\mu\right)\right]^{2}\nonumber\\
&  &  \qquad \times \biggl[U_{11}\left(Q\right)^{2}+U_{12}\left(Q\right)^{2}\biggr]\biggr\}\sum_{\alpha}\cos\left(\phi_{1\alpha}-\phi_{2\alpha}\right),
 \end{eqnarray}
\begin{eqnarray}
\gamma_{TTE}\left(Q,\delta\phi,T\right) 
& = & \biggl\{\frac{1}{2}\biggl[U_{11}\left(Q\right)^{2}U_{12}\left(Q\right)^{2}f'''\left(E_{1}-\mu\right)+U_{12}\left(Q\right)^{2}U_{22}\left(Q\right)^{2}f'''\left(E_{2}-\mu\right)\biggr]\nonumber\\
&    & \qquad-\frac{1}{\sum_{i=1}^{2}f'\left(E_{i}-\mu\right)}
%\nonumber\\ &    & \qquad\times
\left[U_{11}\left(Q\right)U_{12}\left(Q\right)f''\left(E_{1}-\mu\right)+U_{22}\left(Q\right)U_{12}\left(Q\right)f''\left(E_{2}-\mu\right)\right]^{2}\biggr\}\nonumber\\
&    &  \times \sum_{i=1}^{3}\left[1+2\cos^{2}\left(\delta\phi_{i}\right)\right]\nonumber\\
&    & +\biggl\{\frac{2}{3}\biggl[U_{11}\left(Q\right)^{2}U_{12}\left(Q\right)^{2}f'''\left(E_{1}-\mu\right)+U_{12}\left(Q\right)^{2}U_{22}\left(Q\right)^{2}f'''\left(E_{2}-\mu\right)\biggr]\nonumber\\
&    & \qquad\times \left[2+\cos\left(\frac{\sqrt{3}}{2}Q^{2}\right)\right]%\nonumber\\
%&    & \qquad
-\frac{2}{\sum_{i=1}^{2}f'\left(E_{i}-\mu\right)}\nonumber\\
&    & \qquad \times \left[U_{11}\left(Q\right)U_{12}\left(Q\right)f''\left(E_{1}-\mu\right)+U_{22}\left(Q\right)U_{12}\left(Q\right)f''\left(E_{2}-\mu\right)\right]^{2}\biggr\}\nonumber\\
&    & \times \sum_{i=1}^{3}\sum_{j\neq i}\left[1+2\cos\left(\delta\phi_{i}\right)\cos\left(\delta\phi_{j}\right)\right]\nonumber\\
&    & -\frac{9}{\sum_{i=1}^{2}f'\left(E_{i}-\mu\right)}\left[U_{11}\left(Q\right)U_{22}\left(Q\right)-U_{12}\left(Q\right)^{2}\right]^{2}f''\left(E_{1}-\mu\right)f''\left(E_{2}-\mu\right),
% \nonumber\\ &    &
\end{eqnarray}
 and
\begin{eqnarray}
\rho_{TTE}\left(Q,\delta\phi,T\right) 
& = & \biggl\{\frac{1}{3}\left[11+4\cos\left(\frac{\sqrt{3}}{2}Q^{2}\right)\right]%\nonumber\\
%&    & \qquad\times
\left[U_{12}\left(Q\right)^{3}U_{22}\left(Q\right)f'''\left(E_{2}-\mu\right)+U_{11}\left(Q\right)U_{12}\left(Q\right)^{3}f'''\left(E_{1}-\mu\right)\right]\nonumber\\
&    & \qquad-\frac{6}{\sum_{i=1}^{2}f'\left(E_{i}-\mu\right)}
% \nonumber\\&    & \qquad\times
\left[U_{11}\left(Q\right)U_{12}\left(Q\right)f''\left(E_{1}-\mu\right)+U_{22}\left(Q\right)U_{12}\left(Q\right)f''\left(E_{2}-\mu\right)\right]\nonumber\\
&    & \qquad\times\biggl[U_{12}\left(Q\right)^{2}+U_{22}\left(Q\right)^{2}\biggr]\biggr\}\sum_{\alpha}\cos\left(\phi_{1\alpha}-\phi_{2\alpha}\right).
\end{eqnarray}

\subsection{Mixed unidirectional-triangular CDW system}

We consider mixing between a unidirectional phase in $\Phi_{1}$ and
a triangular phase in $\Phi_{2}$. Without loss of generality, we
let $\mathbf{Q}_{1}=\mathbf{Q}_{2}^{1}$. Thus, we end up with two
new parameters: 
\begin{eqnarray}
\delta\phi & = & \phi_{1}-\phi_{2},\\
\tilde{\phi}_{2} & = & \sum_{i=1}^{3}\phi_{2}^{i},
\end{eqnarray}
 The new terms in the free energy theory for STE ordering are

\begin{eqnarray}
\delta f_{STE} & = & 
P_1 P_2 \alpha_{STE}\left(Q,\delta\phi,T\right)\left|\Delta_{1}\right|\left|\Delta_{2}\right|
+ P_1 P_2^2 \varphi_{STE}\left(Q,\delta\phi,\tilde{\phi}_{2},T\right)\left|\Delta_{1}\right|\left|\Delta_{2}\right|^{2}
+ P_1^2 P_2^2 \gamma_{STE}\left(Q,\delta\phi,T\right)\left|\Delta_{1}\right|^{2}\left|\Delta_{2}\right|^{2}
\nonumber \\  & &
+ P_1^3 P_2 \sigma_{STE}\left(Q,\delta\phi,T\right)\left|\Delta_{1}\right|^{3}\Delta_{2} 
+ P_1 P_2^3 \rho_{TTE}\left(Q,\delta\phi,T\right)\left|\Delta_{1}\right|\left|\Delta_{2}\right|^{3}.
\end{eqnarray}
The coefficients are 
\begin{eqnarray}
\alpha_{STE}\left(Q,\delta\phi,T\right) 
& = & 2U_{12}\left(Q\right)\biggl[1+U_{11}\left(Q\right)f'\left(E_{1}-\mu\right)+U_{22}\left(Q\right)f'\left(E_{2}-\mu\right)\biggr]
\cos\left(\delta\phi\right),
\end{eqnarray}

\begin{eqnarray}
\varphi_{STE}\left(Q,\delta\phi,\tilde{\phi}_{2},T\right) 
& = & -2\biggl[ U_{12}\left(Q\right)^{2}U_{11}\left(Q\right)f''\left(E_{1}-\mu\right)+U_{12}\left(Q\right)U_{22}\left(Q\right)^{2}f''\left(E_{2}-\mu\right)\biggr]\nonumber\\
&    & \times\cos\left(\frac{\sqrt{3}}{4}Q^{2}\right)\cos\left(\tilde{\phi}_{2}+\delta\phi\right),
 \end{eqnarray}
 
\begin{eqnarray}
\sigma_{STE}\left(Q,\delta\phi,T\right) 
& = & \biggl\{\frac{1}{3}\left[19+8\cos\left(\frac{\sqrt{3}}{2}Q^{2}\right)\right] %\nonumber\\ &    & \times
\left[U_{11}\left(Q\right)^{3}U_{12}\left(Q\right)f'''\left(E_{1}-\mu\right)+U_{22}\left(Q\right)U_{12}\left(Q\right)^{3}f'''\left(E_{2}-\mu\right)\right]\nonumber\\
&    & \qquad-\frac{1}{\sum_{i=1}^{2}f'\left(E_{i}-\mu\right)} 
%\nonumber\\ &    & \qquad\times
\left[U_{11}\left(Q\right)U_{12}\left(Q\right)f''\left(E_{1}-\mu\right)+U_{22}\left(Q\right)U_{12}\left(Q\right)f''\left(E_{2}-\mu\right)\right]^{2}\nonumber\\
&    & \qquad\times  \biggl(U_{11}\left(Q\right)^{2}+U_{12}\left(Q\right)^{2}\biggr)\biggr\}\cos\left(\delta\phi\right),
 \end{eqnarray}
 
\begin{eqnarray}
\gamma_{STE}\left(Q,\delta\phi,T\right) 
& = & \biggl\{\frac{1}{2}U_{11}\left(Q\right)^{2}U_{12}\left(Q\right)^{2}f'''\left(E_{1}-\mu\right)+U_{12}\left(Q\right)^{2}U_{22}\left(Q\right)^{2}f'''\left(E_{2}-\mu\right)\nonumber\\
&    & \qquad-\frac{1}{\sum_{i=1}^{2}f'\left(E_{i}-\mu\right)}
%\nonumber\\ &    & \qquad\times
\left[U_{11}\left(Q\right)U_{12}\left(Q\right)f''\left(E_{1}-\mu\right)+U_{22}\left(Q\right)U_{12}\left(Q\right)f''\left(E_{2}-\mu\right)\right]^{2}\biggr\}\nonumber\\
&    &\times\left[1+\frac{2}{3}\cos^{2}\left(\delta\phi\right)\right]\nonumber\\
&    &+ \biggl\{\left[3+2\cos\left(\frac{\sqrt{3}}{2}Q^{2}\right)\right]U_{11}\left(Q\right)^{2}U_{12}\left(Q\right)^{2}f'''\left(E_{1}-\mu\right)
%\nonumber\\ &    &\qquad
+U_{12}\left(Q\right)^{2}U_{22}\left(Q\right)^{2}f'''\left(E_{2}-\mu\right)
\nonumber\\
&    & \qquad
-\frac{3}{\sum_{i=1}^{2}f'\left(E_{i}-\mu\right)} 
%\nonumber\\ &    &\qquad\times
\left[U_{11}\left(Q\right)U_{12}\left(Q\right)f''\left(E_{1}-\mu\right)+U_{22}\left(Q\right)U_{12}\left(Q\right)f''\left(E_{2}-\mu\right)\right]^{2}\biggr\}, 
% \nonumber\\ &    &
\end{eqnarray}
and
\begin{eqnarray}
\rho_{STE}\left(Q,\delta\phi,T\right) 
& = & \biggl\{\frac{1}{3}\left[19+8\cos\left(\frac{\sqrt{3}}{2}Q^{2}\right)\right] 
%\nonumber\\ &    & \times
\left[U_{12}\left(Q\right)^{3}U_{22}\left(Q\right)f'''\left(E_{2}-\mu\right)+U_{11}\left(Q\right)U_{12}\left(Q\right)^{3}f'''\left(E_{1}-\mu\right)\right]\nonumber\\
&    & \qquad-\frac{1}{\sum_{i=1}^{2}f'\left(E_{i}-\mu\right)} 
% \nonumber\\ &    & \qquad\times
\left[U_{11}\left(Q\right)U_{12}\left(Q\right)f''\left(E_{1}-\mu\right)+U_{22}\left(Q\right)U_{12}\left(Q\right)f''\left(E_{2}-\mu\right)\right]\nonumber\\
&    & \qquad\times\biggl[U_{12}\left(Q\right)^{2}+U_{22}\left(Q\right)^{2}\biggr]\biggr\}\cos\left(\delta\phi\right).
 \end{eqnarray}

\section{Hartree-Fock potentials for quantum Hall ferromagnets}
\label{sec:HFQHF}

In this appendix, we derive expressions for the potentials $W_{IJ}$ 
in the Hartree-Fock approximation.  We use notation similar to 
Sec.~\ref{sec:Landau-free-energy-theory}, but note that lower-case letters 
label the individual states $\Phi_i$ while upper-case letters 
label the pseudospin indices. 
Let
\begin{eqnarray}
F_{ij}\left(\mathbf{q}\right) & = & \sum_{\alpha} C_{i\alpha}^* C_{j\alpha} e^{-i\left(n_{i\alpha}-n_{j\alpha}\right)\left(\theta-\pi/2\right)}A_{n_{i\alpha}n_{j\alpha}}\left[\frac{\left(ql_0\right)^2}{2}\right],
\end{eqnarray}
where 
\begin{eqnarray*}
A_{mn}\left(x\right) & = & \sqrt{\frac{m!}{n!}} x^{\frac{n-m}{2}}\mathcal{L}_m^{n-m}\left(x\right),\mbox{ }m\le n.
\end{eqnarray*}
Then
\begin{eqnarray*}
\rho\left(\mathbf{q}\right) & = & e^{-\frac{\left(ql_0\right)^2}{4}}\sum_{ij=1}^{2}F_{ij}\left(\mathbf{q}\right)\sum_X e^{-iq_xX} a_{iX_{+}}^\dagger a_{jX_{-}}
\end{eqnarray*}
We invert  Eq.~(\ref{eq:QHFoperators}) to solve for $a_{iX_{+}}^\dagger a_{jX_{-}}$ in terms of the $\hat{m}_{I,\mathbf{q}}$ operators and rewrite $\rho\left(\mathbf{q}\right)$ as follows:
\begin{eqnarray*}
\rho\left(\mathbf{q}\right) & = & e^{-\frac{\left(ql_0\right)^2}{4}}\sum_{IJ=0,x,y,z}\mathcal{F}_{I}\left(\mathbf{q}\right)\hat{m}_{I,\mathbf{q}},
\end{eqnarray*}
where $\mathcal{F}_{I}\left(\mathbf{q}\right) = \frac{1}{2}\sigma_I^{\left(ij\right)}F_{ij}\left(\mathbf{q}\right)$. The density-density Hamiltonian is then

\begin{eqnarray}
\label{eq:H_QHF}
H_{QHF} & = & \frac{1}{2}\sum_{\mathbf{q}}\sum_{IJ}e^{-\frac{\left(ql_0\right)^2}{2}}{\fx {V_{IJ}} \bq} \hat{m}_{I,\mathbf{q}}\hat{m}_{J,-\mathbf{q}} ,
\end{eqnarray}
where ${\fx {V_{IJ}} \bq} = {\fx v \bq}{\fx {\mathcal{F}_I} \bq} {\fx {\mathcal{F}_j} {-\bq}}$, ${\fx v \bq}$ being the electron-electron interaction potential.

We apply the HF approximation to Eq.~(\ref{eq:H_QHF}) to obtain 
Eq.~(\ref{eq:QHFHamiltonian}). The HF potentials $W_{IJ}$ are defined as follows:
\begin{eqnarray*}
{W_{00}} & = & \frac{1}{2\pi}\int{\rm d}^2\bq  \, e^{-\frac{\left(ql_0\right)^2}{2}}\left[{\fx {V_{00}} \bq} - {\fx {V_{00}} 0} -\frac{1}{2}{\fx u \bq}\right],\\
{W_{II}}   & = & \frac{1}{2\pi}\int{\rm d}^2\bq \, e^{-\frac{\left(ql_0\right)^2}{2}}\left[{\fx {V_{II}} \bq} - {\fx {V_{II}} 0} +\frac{1}{2}{\fx u \bq}\right],\\
{W_{IJ}}  & = & \frac{1}{2\pi}\int{\rm d}^2\bq \, e^{-\frac{\left(ql_0\right)^2}{2}}\left[{\fx {V_{IJ}} \bq} - {\fx {V_{IJ}} 0} \right],\\
\end{eqnarray*}
where
\begin{eqnarray*}
{\fx {u} \bq}           & = & {\fx {F_{11}} \bq}{\fx {F_{22}} \bq}-{\fx {F_{12}} \bq}{\fx {F_{21}} \bq}.
\end{eqnarray*}
For each potential, conservation of LL index in the basis state interactions must be observed. Thus, for the eigenstates considered in Sec.~\ref{sec:Numerics}, the only nonzero potentials are $W_{00}$, $W_{zz}$, $W_{0z} = W_{z0}$, and $W_{xx}=W_{yy}$.

\end{document}